\documentclass[sn-nature]{sn-jnl}% Style for submissions to Nature Portfolio journals
\pdfoutput=1
\usepackage{graphicx}%
\usepackage{multirow}%
\usepackage{amsmath,amssymb,amsfonts}%
\usepackage{amsthm}%
\usepackage{mathrsfs}%
\usepackage[title]{appendix}%
\usepackage{xcolor}%
\usepackage{textcomp}%
\usepackage{manyfoot}%
\usepackage{booktabs}%
\usepackage{algorithm}%
\usepackage{algorithmicx}%
\usepackage{algpseudocode}%
\usepackage{listings}%

\begin{document}

\title[Article Title]{SBSM-Pro: Support Bio-sequence Machine for Proteins}

\author[1,2]{\fnm{Yizheng} \sur{Wang}}\email{wyz020@126.com}

\author[1,2]{\fnm{Yixiao} \sur{Zhai}}\email{zhai1xiao@std.uestc.edu.cn}

\author*[2]{\fnm{Yijie} \sur{Ding}}\email{wuxi\_dyj@163.com}

\author*[1,2]{\fnm{Quan} \sur{Zou}}\email{zouquan@nclab.net}

\affil[1]{\orgdiv{Institute of Fundamental and Frontier Sciences}, \orgname{University of Electronic Science and Technology of China}, \orgaddress{\city{Chengdu}, \postcode{611731}, \state{Sichuan}, \country{China}}}

\affil[2]{\orgdiv{Yangtze Delta Region Institute (Quzhou)}, \orgname{University of Electronic Science and Technology of China}, \orgaddress{\city{Quzhou}, \postcode{324003}, \state{Zhejiang}, \country{China}}}

\abstract{Proteins play a pivotal role in biological systems. The use of machine learning algorithms for protein classification can assist and even guide biological experiments, offering crucial insights for biotechnological applications. We introduce the Support Bio-Sequence Machine for Proteins (SBSM-Pro), a model purpose-built for the classification of biological sequences. This model starts with raw sequences and groups amino acids based on their physicochemical properties. It incorporates sequence alignment to measure the similarities between proteins and uses a novel multiple kernel learning (MKL) approach to integrate various types of information, utilizing support vector machines for classification prediction. The results indicate that our model demonstrates commendable performance across ten datasets in terms of the identification of protein function and posttranslational modification. This research not only exemplifies state-of-the-art work in protein classification but also paves avenues for new directions in this domain, representing a beneficial endeavor in the development of platforms tailored for the classification of biological sequences. SBSM-Pro is available for access at http://lab.malab.cn/soft/SBSM-Pro/.}


\maketitle

\section{Introduction}\label{Introduction}

Bio-sequences, which include DNA, RNA, and proteins, are the molecular foundation of modern genetic research. The classification of bio-sequences based on sequence information has been a key focus in bioinformatics research. At present, with the sequential completion of genome mapping from humans to various species, we have amassed a vast amount of sequence data, creating an urgent need for computer-assisted annotation of sequence functions. Although it is statistically evident that genetic sequences determine hereditary diseases, the mechanisms by which sequence variations contribute to diseases are intricately complex. It is difficult to address and interpret all these issues through one biological experiment; hence, multiple computer predictions are needed to guide the progression of wet lab exploration. In summary, the application of information science and machine learning to bio-sequence classification is a valuable tool for assisting researchers in comprehending and analysing bio-sequences. It serves as a key driving force for advancing research in the field of bioinformatics.

In the field of bio-sequence classification, machine learning methods are broadly pursued using two strategies: feature extraction combined with traditional classification methods and direct sequence classification via deep learning techniques.

For bio-sequences, relevant features are mainly characterized as frequency, physicochemical, structural, and evolutionary features. Several notable tools for sequence feature extraction include PseKNC-General \cite{PseKNC_General}, PyFeat \cite{PyFeat}, iFeature \cite{iFeature}, VisFeature \cite{VisFeature}, POSSUM \cite{POSSUM}, Rcpi \cite{Rcpi}, and protr \cite{protr}. Furthermore, every alphabet in the sequence, whether amino acids or nucleotides, can be numerically represented, thereby contributing to the global feature of the sequence \cite{DiProDB, AAindex}.

Given these traditional numerical classification features, classifiers can be integrated to facilitate the classification and discrimination of biological sequences. This led to the emergence of platforms that combine feature extraction and classifiers, such as gkmSVM \cite{gkmSVM}, iLearnPlus \cite{iLearnPlus}, Biological Seq-Analysis2.0 \cite{BioSeq_Analysis2.0}, and BioSeq-BLM \cite{BioSeq-BLM}. Notably, gkmSVM was one of the first to use kernel methods for biological sequence predictions, with the most common frequency feature being k-mer, and yielded promising results in certain scenarios, such as predicting enhancer activity in specific cell types \cite{10.1371/journal.pcbi.1003711} and disease-relevant mutations \cite{lee2015method}. However, the performance of gkmSVM frequently falls short due to its exclusive reliance on rudimentary k-mer features and its susceptibility to overfitting. Both iLearnPlus and Biological Seq-Analysis 2.0 offer a rich array of feature extraction and analysis methods, making them more commonly employed in biological sequence classification research compared to traditional tools. However, these tools do not account for sequence structural information. The recently developed BioSeq-BLM platform offers numerous biological language models for the automated representation and analysis of biological sequence data, enabling the extraction of latent semantic features of biological sequences.

Deep learning-based methods circumvent the need for feature extraction by directly encoding sequences into neural networks. Through training, the architecture and parameters of the network are fine-tuned, enabling it to classify the training samples effectively. The most renowned application of this approach is AlphaFold2’s \cite{AlphaFold} prediction of protein 3D structures, facilitated by the advent of cryo-electron microscopy, which provides a wealth of 3D structural samples for AI training. Platforms such as Kipoi \cite{Kipoi}, Pysster \cite{pysster}, Selene \cite{Selene}, and DNA-BERT \cite{DNABERT} have been developed for deep learning-based classification of biological sequences. Autoencoders, a type of artificial neural network, are used to learn effective data encoding in an unsupervised manner. For instance, autoencoders have been utilized to analyze over one thousand yeast microarray datasets, facilitating the exploration of the yeast transcriptional regulatory code \cite{chen2016learning}. Research has shown hidden variables in the first layer capture signals of yeast transcription factors (TFs) effectively, establishing a nearly one-to-one mapping between the hidden variables and TFs. Inspired by biological processes, convolutional neural networks (CNNs), whose connectivity patterns between neurons resemble those of the animal visual cortex, are commonly used for various sequence data to learn the inherent regularities \cite{10.1093/bioinformatics/btw427, 10.1093/bioinformatics/btw255, 10.1093/nar/gkx177, 7822593, 10.1093/bioinformatics/bty228, zhou2015predicting} and specificities \cite{alipanahi2015predicting} within gene sequences. By reincorporating newly discovered sequence motifs into the neural network model and continuously updating the model’s predictive scores, accuracy in predicting sequence specificity can be improved, enabling the analysis of potentially pathogenic genomic variations. Recurrent neural networks (RNNs) accumulate sequence information over time. Hybrid predictive models combining both CNNs and RNNs are currently popular and have been applied in various computational biology domains, including DNA methylation \cite{angermueller2017deepcpg}, chromatin accessibility \cite{10.1093/bioinformatics/btx234}, and noncoding RNAs \cite{10.1093/nar/gkw226}. CNN layers are adept at capturing prevalent regulatory motifs, while RNN layers excel at capturing the enduring dependencies among these motifs, facilitating the learning of “syntax” rules to improve the prediction performance.

It is anticipated that machine learning methods will continue to prosper in future biological sequence research. This trend is facilitated by significant advancements in methodologies, software, and hardware. Researchers are also striving to promote biological studies by innovating machine learning strategies. Historically, the majority of achievements in this domain have been realized through the adaptation and direct application of algorithms initially developed in other fields to biological data. Classical CNNs and RNNs, as well as more recently celebrated transformer architectures like BERT and GPT, have their origins in domains like image analysis (for tasks such as face recognition or autonomous driving) and natural language processing. However, there is no universal algorithm or framework specifically tailored for biological sequence data. The development of custom algorithms specifically designed for these types of data and problems represents one of the most exciting prospects in bioinformatics. In the realm of biological sequence classification, the three paramount issues revolve around the universality of methods, the user-friendliness of software, and the accuracy of prediction. Traditional alignment-based methods fall short in terms of universality due to their inherent inability to negate the impact of nonfunctional sequence intervals, necessitating the integrated use of machine learning methodologies. Conventional machine learning approaches, despite their advantage in facilitating the development of user-friendly predictive software platforms through numerical feature extraction, have limitations. Relying solely on word frequency, physicochemical, and evolutionary features falls short in capturing the full spectrum of sequence information, thus invariably imposing a ceiling on prediction accuracy. On the other hand, deep learning-based approaches pose unique challenges. They demand a substantial volume of training data to avoid overfitting, and the complexity of deep learning software packages can detract from the user-friendliness of the sequence classification platform. These complexities may discourage researchers lacking a background in information science from utilizing these tools.

In response to the challenges encountered by the aforementioned research approaches, building on the strengths of support vector machines (SVMs), especially their prowess in effectively managing small sample problems, we present an innovative methodology termed the support bio-sequence machine for proteins (SBSM-Pro). This method takes a unique approach by replacing numerical vectors with biological sequences, harnessing the power of sequence alignment algorithms. In doing so, it eliminates the need for deep learning's reliance on extensive data volumes. We establish an end-to-end kernel method from sequence to metric. SBSM-Pro is applied to predict the structure and function of biological sequences. Concurrently, it effectively mines the underlying patterns and feature interpretability of adaptive variable-length sequence fragments.

In this paper, we make several key contributions to the field, which are given as follows: (i) We propose a novel standard process named physicochemical properties-spectral clustering-dictionaries (PSD) that effectively reduces the amino acid alphabet. This process facilitates sequence alignment and accurately represents the distances between proteins, thereby linking the physicochemical properties of proteins with their sequences. (ii) We introduce two methods for calculating sequence similarity kernels, namely, the Levenshtein (LS) distance and the Smith‒Waterman (SW) score. These techniques allow for precise comparisons between protein sequences. (iii) We present a new multiple kernel learning (MKL) approach that combines global and local kernels, thus effectively integrating multiple similarity kernels. This distinctive method optimizes the processing and understanding of protein data. (iv) We employ an SVM with a precomputed kernel to receive the fused sequence kernels for protein prediction. This machine learning model ensures efficient and precise prediction. These combined contributions present a comprehensive and innovative approach to the analysis and prediction of protein sequences.

\section{Results and Discussion}\label{Results}

\subsection{Performance metric}

We employed accuracy (ACC), a widely recognized and indispensable performance metric for classification models. Given that existing methods adopt ACC as their performance metric, we chose the same criterion to facilitate a more direct comparison. The formula for calculating ACC is as follows:

\begin{equation}
    ACC = \frac{{TP + TN}}{{TP + TN + FP + FN}}
\end{equation}
where TP, TN, FN, and FP denote the number of true positives, true negatives, false negatives, and false positives, respectively.

\subsection{Comparative Analysis of the Proposed Method and Existing Methods}

\begin{figure}[!ht]%
\centering
\includegraphics[width=0.7\textwidth]{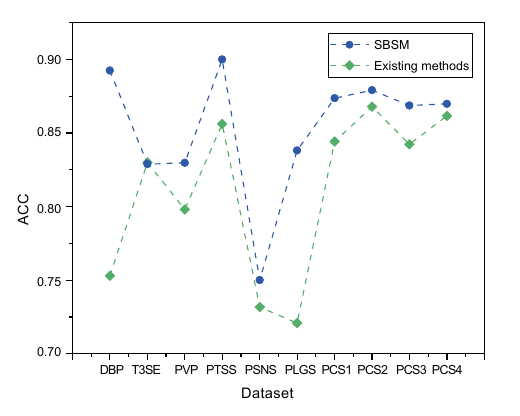}
\caption{\textbf{Line graph comparison of ACC values between the proposed method and the existing methods.} Distinct colored lines represent SBSM-Pro and existing methods.}\label{Existing methods}
\end{figure}

To achieve a significant breakthrough with SBSM-Pro, we compared it with the leading contemporary models to evaluate its effectiveness. To demonstrate the robustness of SBSM-Pro, we selected ten commonly used protein classification datasets. The results are shown in Table \ref{tab1} and Figure \ref{Existing methods}.

\begin{table}[h]
\centering
\begin{tabular}{ccc}
\hline
& SBSM-Pro & Existing methods \\
\hline
DBP & \textbf{0.8925} & 0.753 \\
T3SE & 0.8289 & \textbf{0.83} \\
PVP & \textbf{0.8298} & 0.798 \\
PTSS & \textbf{0.9000} & 0.8563 \\
PSNS & \textbf{0.7500} & 0.7317 \\
PLGS & \textbf{0.8381} & 0.7207 \\
PCS1 & \textbf{0.8737} & 0.8443 \\
PCS2 & \textbf{0.8791} & 0.8679 \\
PCS3 & \textbf{0.8687} & 0.8423 \\
PCS4 & \textbf{0.8699} & 0.8617 \\
\hline
\end{tabular}
\caption{\textbf{Comparison of ACC values between the proposed method and existing methods.} Each row displays a comparison of the ACC values between our model, SBSM-Pro, and the best existing methods across ten datasets. The highest value in each row is highlighted in bold.}
\label{tab1}
\end{table}

The previously established method for the DBP dataset, introduced by Lu et al. \cite{Dataset1}, is a model founded on SVM. This model extracts evolutionary features and concatenates them as input for the model. However, due to its exclusive emphasis on evolutionary features, this model overlooks certain information. As a result, the ACC of SBSM-Pro surpasses that of the aforementioned method by 0.1853.

Hui et al. \cite{Dataset2} developed the T3SEpp model, which exhibits the best performance with the T3SE dataset. This model integrates both traditional machine learning models, such as SVM and random forests, and deep learning models, such as fully connected neural networks and convolutional neural networks. The performance of SBSM-Pro is on par with that of T3SEpp. It is important to note that due to numerical precision differences resulting from retaining significant figures, the performance of SBSM-Pro is not necessarily inferior to that of T3SEpp.

For the PVP and PTSS datasets, the models proposed by Meng et al. \cite{Dataset3} and Barukab et al. \cite{Dataset4} are currently the best. Both models primarily utilize amino acid composition information, with the former additionally incorporating feature selection algorithms. However, it's worth noting that both models suffer from overfitting issues.. In terms of ACC, the performance of SBSM-Pro surpasses that of these methods by approximately 3.98\% and 5.10\%, respectively.

Li et al. \cite{Dataset5} employed a set of nine features, including the parallel correlation pseudo amino acid composition and adapted normal distribution bi-profile Bayes, to identify PSNS. This model accounts for a rich set of information, subsequently employing the method of information gain for feature vector selection. However, this approach leads to information loss during feature extraction and selection. Consequently, when using the original protein sequence, SBSM-Pro continues to outperform the traditional numerical vector-based method, showing a 2.50\% improvement in ACC.

Dou et al. \cite{Dataset6} developed iGlu\_AdaBoost, a tool designed for the identification of PLGS. This model integrates three feature representation methods: a 188-dimensional feature, the position of K-spaced amino acid pairs, and the enhanced amino acid composition. By applying feature selection, a 37-dimensional optimal feature subset was obtained, and predictions were performed using AdaBoost. However, there is potential for enhancing the model's generalizability. SBSM-Pro achieves an accuracy (ACC) that surpasses iGlu\_AdaBoost by 16.29\%.

The iCar-PseCp \cite{Dataset7to10} is utilized for the identification of PCSs, employing sequence coupling effects to describe the sequence order with the aim of preserving more information from the original sequence. However, its performance remains less impressive than that of SBSM-Pro when using the original sequence directly. Across the four PCS datasets, SBSM-Pro’s ACC is better by 3.48\%, 1.30\%, 3.13\%, and 0.95\%, respectively.

In summary, when compared with SBSM-Pro, some models incorporate features from multiple dimensions, encompassing a wealth of information. However, they still inevitably suffer from information loss during feature extraction and selection. On the other hand, other models employ deep learning techniques, but this leads to overfitting. Across 10 commonly used amino acid classification datasets, SBSM-Pro generally outperforms existing methods, effectively demonstrating its superior performance, generalizability, and robustness.

\subsection{Creating Dictionaries for Amino Acid Grouping by Using Spectral Clustering}

\begin{figure}[!ht]%
\centering
\includegraphics[width=1\textwidth]{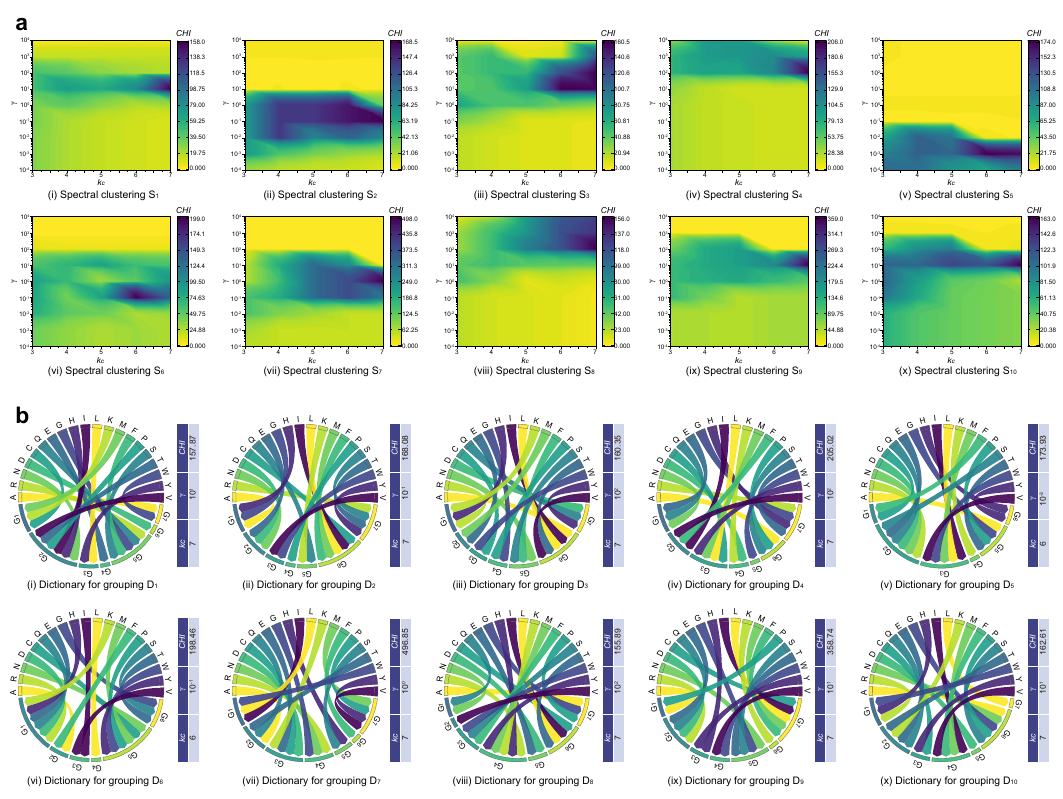}
\caption{\textbf{Overview of the PSD process.} \textbf{a}, Heatmaps of grid search parameter tuning in spectral clustering. Color mapping represents the magnitude of the CHI values, followed by generating a continuous color image to illustrate how CHI varies with changes in $k_{c}$ and $\gamma$. Darker colors indicate higher CHI values, suggesting a better combination of parameters. \textbf{b}, Visual representations of dictionaries for grouping. The upper half of the circle depicts 20 common amino acids, while the lower half showcases specific groups of amino acids. The amino acids in the upper section are linked to their corresponding groups below by arrows, signifying their affiliation. Adjacent to the circle on the right is a table detailing the parameters $k_{c}$ and $\gamma$ used for spectral clustering of the given groups, along with their respective CHI values. }\label{PSD}
\end{figure}

In the process of spectral clustering, we employed a grid search to adjust the hyperparameters. The results of the grid search are shown in Figure \ref{PSD}(a).

The hyperparameter $\gamma$ in the Gaussian kernel function need to be specified by the user prior to using the algorithm. The range of hyperparameter $\gamma$ is on a logarithmic scale from $10^{-4}$ to $10^{4}$, and we used a geometric sequence with a common ratio of $10$. Thus, we performed a grid search within the range $[10^{-4}, 10^{-3}, \cdots, 10^{4}]$.

-means is an unsupervised learning algorithm that necessitates the pre-determination of the number of clusters through the hyperparameter $k_{c}$. A small number of clusters may lead to a significant loss of original sequence information, while an overly large number of clusters may fail to effectively reduce the amino acid alphabet. Consequently, the range for the hyperparameter $k_{c}$ is 3 to 7, with a step size of 1, leading to a grid search within the range $[3, 4, \cdots, 7]$.

The possible values for the hyperparameters $k_{c}$ and $\gamma$ constitute a parameter grid. For each combination of parameters, we trained a spectral clustering model and evaluated the performance of the clustering results. We employed CHI as our evaluation metric and selected the combination of parameters that maximizes this metric as our final configuration for the hyperparameters. For 10 different physicochemical properties of amino acids, we obtained 10 corresponding spectral clustering results. Then, we obtained 10 dictionaries for grouping based on 10 different clustering results, completing the PSD process, as shown in Tables \ref{D1} to \ref{D10} in Appendix \ref{Supplementary material}. Their visual representations are depicted in Figure \ref{PSD}(b). According to the clustering results, we found that the number of clusters in dictionaries $D_5$ and $D_6$ is six, whereas the remaining eight dictionaries each comprise seven clusters. Each group in the dictionaries contains a maximum of 6 amino acids and a minimum of 1 amino acid.

\subsection{Comparison of the Effect of Different Dictionaries for Amino Acid Grouping}

In the previous section, we obtained ten dictionaries for amino acid grouping. Based on one of these dictionaries, the amino acid residues of the original protein sequence were replaced by the identification number of their respective groups, resulting in the re-encoding of the protein.

\begin{figure*}[!ht]%
\centering
\includegraphics[width=1\textwidth]{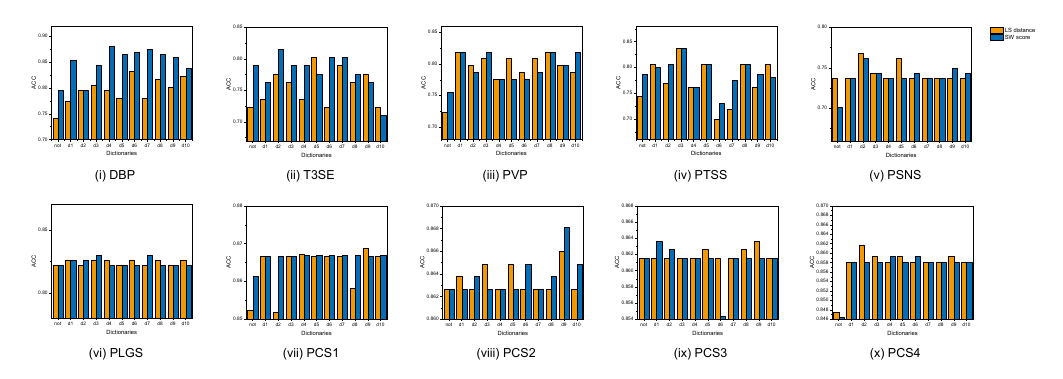}
\caption{\textbf{Bar chart to compare the effects of different dictionaries for grouping.} The orange and blue bars to represent the LS distance and SW score, respectively. We compared the performance of models with 10 datasets, distinguishing between those that utilize a dictionary and those that do not, with the latter being labelled as "not".}\label{dictionaries}
\end{figure*}

This process can reduce interference in the sequence alignment, while also linking the original amino acid sequence information to its physicochemical properties. As a result, it effectively enhances the efficiency of LS distance and SW scores in quantifying protein similarity. To substantiate this perspective and highlight the role of the PSD process, we compared the results after substitution with different dictionaries for amino acid grouping to those without any amino acid substitution. Two methods for measuring the amino acid similarity, the LS distance and SW score, were employed to compare the results of dictionaries for grouping. The specific results are presented in Table \ref{LS} and Table \ref{SW}, respectively, while the overall outcomes are illustrated in Figure \ref{dictionaries}.

\begin{table}[!ht]
\centering
\begin{tabular}{lllllllllll}
\toprule
 & DBP & T3SE & PVP & PTSS & PSNS & PLGS & PCS1 & PCS2 & PCS3 & PCS4 \\ 
\midrule
not & 0.7419 & 0.7368 & 0.8085 & 0.7438 & 0.7378 & 0.8219 & 0.8667 & 0.8627 & 0.8616 & 0.8476 \\
d1  & 0.7742 & 0.7368 & 0.8191 & 0.8063 & 0.7378 & 0.8259 & 0.8667 & 0.8638 & 0.8616 & 0.8581 \\
d2  & 0.7957 & 0.7763 & 0.7979 & 0.7688 & 0.7683 & 0.8219 & 0.8667 & 0.8627 & 0.8616 & 0.8617 \\
d3  & 0.8064 & 0.7632 & 0.8085 & 0.8375 & 0.7439 & 0.8259 & 0.8667 & 0.8649 & 0.8616 & 0.8593 \\
d4  & 0.7957 & 0.7368 & 0.7766 & 0.7625 & 0.7378 & 0.8259 & 0.8671 & 0.8627 & 0.8616 & 0.8581 \\
d5  & 0.7796 & 0.8026 & 0.8085 & 0.8063 & 0.7622 & 0.8219 & 0.8667 & 0.8649 & 0.8626 & 0.8593 \\
d6  & 0.8333 & 0.7237 & 0.7872 & 0.7000 & 0.7378 & 0.8259 & 0.8667 & 0.8627 & 0.8616 & 0.8581 \\
d7  & 0.7796 & 0.7895 & 0.8085 & 0.7188 & 0.7378 & 0.8219 & 0.8667 & 0.8627 & 0.8616 & 0.8581 \\
d8  & 0.8172 & 0.7632 & 0.8191 & 0.8063 & 0.7378 & 0.8259 & 0.8667 & 0.8627 & 0.8626 & 0.8581 \\
d9  & 0.8011 & 0.7763 & 0.7979 & 0.7625 & 0.7378 & 0.8219 & 0.8688 & 0.8660 & 0.8636 & 0.8593 \\
d10 & 0.8226 & 0.7237 & 0.7872 & 0.8063 & 0.7378 & 0.8259 & 0.8667 & 0.8627 & 0.8616 & 0.8581 \\
\bottomrule
\end{tabular}
\caption{\textbf{Comparison of the effect of different dictionaries for grouping by the LS distance.} The columns in the table represent the performance with specific datasets. Models that do not use dictionaries for grouping (denoted as "not") and those using 10 different dictionaries are included in this table.}
\label{LS}
\end{table}

\begin{table}[!ht]
\centering
\begin{tabular}{lllllllllll}
\toprule
 & DBP & T3SE & PVP & PTSS & PSNS & PLGS & PCS1 & PCS2 & PCS3 & PCS4 \\ 
\midrule
not & 0.7957 & 0.7237 & 0.7234 & 0.7875 & 0.7012 & 0.8219 & 0.8524 & 0.8627 & 0.8616 & 0.8464 \\
d1  & 0.8548 & 0.7632 & 0.8191 & 0.8000 & 0.7378 & 0.8259 & 0.8667 & 0.8627 & 0.8636 & 0.8581 \\
d2  & 0.7957 & 0.8158 & 0.7872 & 0.8063 & 0.7622 & 0.8259 & 0.8519 & 0.8638 & 0.8626 & 0.8581 \\
d3  & 0.8441 & 0.7895 & 0.8191 & 0.8375 & 0.7439 & 0.8300 & 0.8667 & 0.8627 & 0.8616 & 0.8581 \\
d4  & 0.8817 & 0.7895 & 0.7766 & 0.7625 & 0.7378 & 0.8219 & 0.8670 & 0.8627 & 0.8616 & 0.8593 \\
d5  & 0.8656 & 0.7763 & 0.7766 & 0.8063 & 0.7378 & 0.8219 & 0.8670 & 0.8627 & 0.8616 & 0.8581 \\
d6  & 0.8701 & 0.8026 & 0.7766 & 0.7313 & 0.7439 & 0.8219 & 0.8670 & 0.8649 & 0.8544 & 0.8593 \\
d7  & 0.8763 & 0.8026 & 0.7872 & 0.7750 & 0.7378 & 0.8300 & 0.8582 & 0.8627 & 0.8616 & 0.8581 \\
d8  & 0.8656 & 0.7763 & 0.8191 & 0.8063 & 0.7378 & 0.8219 & 0.8670 & 0.8638 & 0.8616 & 0.8581 \\
d9  & 0.8602 & 0.7632 & 0.7979 & 0.7875 & 0.7500 & 0.8219 & 0.8667 & 0.8681 & 0.8616 & 0.8581 \\
d10 & 0.8387 & 0.7105 & 0.8191 & 0.7813 & 0.7439 & 0.8219 & 0.8670 & 0.8649 & 0.8616 & 0.8581 \\
\bottomrule
\end{tabular}
\caption{\textbf{Comparison of the effect of different dictionaries for grouping by the SW score.}  The columns in the table represent the performance with specific datasets. Models that do not use dictionaries for grouping (denoted as "not") and those using 10 different dictionaries are included in this table.}
\label{SW}
\end{table}

The results indicate that models utilizing amino acid grouping generally outperform those that do not incorporate this grouping. These results align with our expectations and demonstrate the intended benefits of amino acid grouping. For specific datasets, such as T3SE, some models exhibited lower performance when using dictionaries compared to not using them.

The results indicate that for the functional protein classification datasets DBP, T3SE, and PVP, the models based on amino acid grouping achieved significant improvements in ACC compared to those without amino acid grouping. For the other seven PTM identification datasets, the effect of amino acid grouping enhancement was not as pronounced. Referring to Figure \ref{boxplots}, we attribute this observation to the relatively shorter protein sequences in the PTM datasets compared to those in the amino acid function identification datasets. This, however, does not fully demonstrate the effect of reducing the amino acid alphabet, thus diminishing the noise reduction benefits of amino acid grouping. Additionally, overall, the SW algorithm consistently outperforms the LS algorithm. This advantage is due to the SW algorithm’s ability to insert gaps during sequence alignment, resulting in better sequence alignments and, consequently, a more accurate representation of sequence similarity.

The results also r unveiled another noteworthy observation: the performance of different dictionaries varies across datasets. For instance, when using the LS distance, amino acid groupings based on dictionary $d_5$, which includes protein secondary structure information, achieved the highest performance with the DBP dataset but performed the lowest with T3SE. The reason for this discrepancy is that different PSD processes produce dictionaries corresponding to different physicochemical properties of amino acids, and the contributions of these properties to protein classification vary among datasets.

In conclusion, the use of amino acid grouping can provide substantial performance improvements. However, no single amino acid dictionary exhibits good performance across all datasets. This introduces another challenge: the crucial task of selecting the most suitable dictionary. We innovatively addressed this concern by utilizing MKL to integrate similarity kernels generated from all dictionaries. Different kernels are assigned varying weights, leveraging the potential of each amino acid dictionary. This method will be elucidated in the following section.

\subsection{Multiple kernel learning}

In the previous section, we derived dictionaries for amino acid groupings, each corresponding to distinct physicochemical properties. For each dictionary, we selected two distinct sequence similarity measurement methods: the LS distance and SW scores. These two methods offer different perspectives when assessing protein sequence similarity. Consequently, by integrating 10 amino acid substitution dictionaries with the 2 sequence similarity measurement techniques, we obtained 20 protein similarity kernels. These 20 kernels represent a multidimensional evaluation of protein sequence similarity, with each kernel having a unique characterization capability.

Utilizing hybrid central kernel dependence maximization MKL (HCKDM-MKL), we obtained weights for the 20 similarity kernels. These weights signify the contribution of each similarity kernel in the fused kernel. To visually represent the weight of each similarity kernel as well as the proportions of contributions from LS distance and SW scores, we constructed a concentric ring chart, as shown in Figure 4. Examining the kernel weight figures allows us to summarize various typical patterns of kernel weights obtained through the HCKDM-MKL method. The first type, represented by T3SE, effectively utilizes all 20 similarity kernels. These kernels control the importance of different information with weights. The second type utilizes only a few or even a special similarity kernel, exemplified by PSNS. Regardless of the type, they both employ the HCKDM-MKL method to select relevant information, and their effectiveness has been demonstrated in experiments.

\begin{figure*}[!ht]%
\centering
\includegraphics[width=1\textwidth]{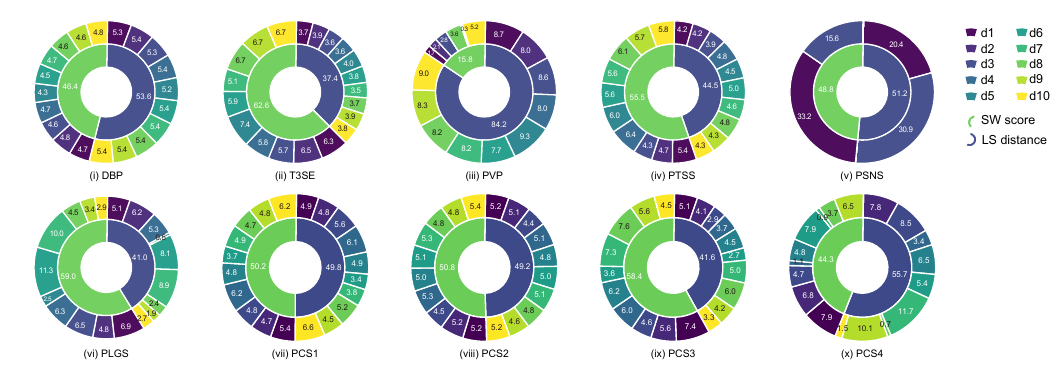}
\caption{\textbf{Concentric ring diagram illustrating the proportional kernel weights computed by HCKDM-MKL.} The inner ring of the circle represents the proportions of two similarity measurement methods, the LS distance and SW score methods, each of which corresponds to the dictionaries depicted by different colors in the outer ring. The combination of ten dictionaries with two measurement methods results in a total of 20 similarity kernels. The weight proportions of these kernels within the fused kernel are visually represented in the outer ring.}\label{kernel weights}
\end{figure*}

\begin{figure*}[!ht]%
\centering
\includegraphics[width=1\textwidth]{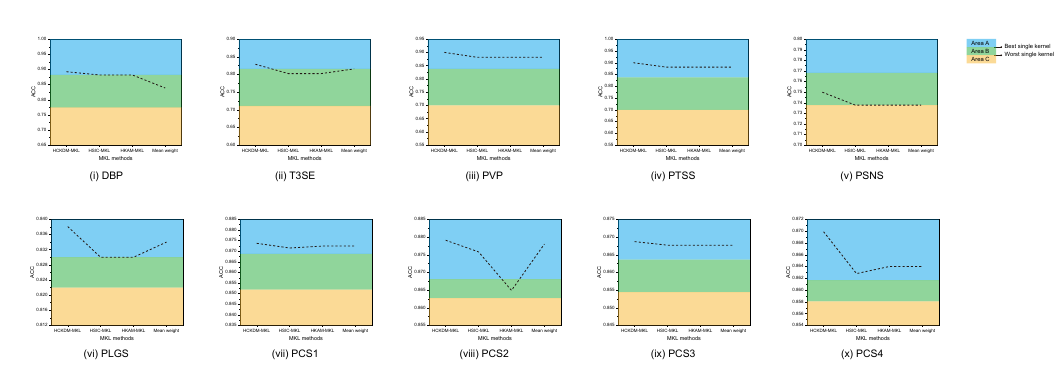}
\caption{\textbf{Line graph comparing the effectiveness of different MKL methods.} T The two lines depict the performance of the top-performing and least-performing kernels. These lines divide the chart into three sections, colored blue, green, and yellow, corresponding to areas A, B, and C, respectively. In area A, the MKL approach demonstrates superior performance. Through MKL, not only are weights assigned to different kernel matrices, highlighting the importance of well-performing kernels, but the performance is further enhanced, surpassing that of any single kernel. The effective method in area B only accomplishes the function of kernel selection, while the MKL method appearing in area C is considered substandard.}\label{MKL}
\end{figure*}

To highlight the effectiveness of our newly proposed MKL method, HCKDM-MKL, we also compared it with two common MKL methods: Hilbert‒Schmidt independence criterion MKL (HSIC-MKL) \cite{hsic} and hybrid kernel alignment maximization MKL (HKAM-MKL) \cite{hkam}. Furthermore, we included the commonly-used method of average kernel weights for comparison, aiming to evaluate the efficacy of MKL approaches. The results are presented in Table \ref{tab3} and Figure \ref{MKL}. We found that the performance of HCKDM-MKL consistently surpassed that of HSIC-MKL and HKAM-MKL in terms of the mean weight across all datasets. This underscores the advanced nature and robustness of our method. In Figure \ref{MKL}, two lines representing the best and worst performance of a single kernel are used to divide the graph into three areas, labelled A, B, and C. The MKL method in area A implies that its performance surpasses that achieved with all kernels, representing the optimal scenario. The methods in area B solely achieve the task of kernel selection. In contrast, the method found in area C is deemed to be below par. We observe that, apart from the PSNS dataset, HCKDM-MKL consistently falls within area A. This suggests that it effectively accomplishes kernel fusion by appropriately assigning weights to different kernels. This even leads to a notable enhancement in the final results, which aligns perfectly with our expectations.

\begin{table}[!ht]
\centering
\begin{tabular}{ccccccc}
\hline
 & HCKDM-MKL & HSIC-MKL & HKAM-MKL & Mean weight \\
\hline
DBP  & \textbf{0.8925} & 0.8817 & 0.8817 & 0.8387 \\
T3SE & \textbf{0.8289} & 0.8026 & 0.8026 & 0.8158 \\
PVP  & \textbf{0.8298} & 0.8085 & 0.8191 & 0.8298 \\
PTSS & \textbf{0.9000} & 0.8813 & 0.8813 & 0.8813 \\
PSNS & \textbf{0.7500} & 0.7378 & 0.7378 & 0.7378 \\
PLGS & \textbf{0.8381} & 0.8300 & 0.8300 & 0.8340 \\
PCS1 & \textbf{0.8737} & 0.8715 & 0.8724 & 0.8724 \\
PCS2 & \textbf{0.8791} & 0.8758 & 0.8649 & 0.8780 \\
PCS3 & \textbf{0.8687} & 0.8677 & 0.8677 & 0.8677 \\
PCS4 & \textbf{0.8699} & 0.8628 & 0.8640 & 0.8640 \\
\hline
\end{tabular}
\caption{\textbf{Comparison of the effects of different MKL methods.} EEach column represents a different MKL method. Each row illustrates the performance of various methods with different datasets. The highest values are in bold.}
\label{tab3}
\end{table}

\section{Conclusion}\label{Conclusion}

The SBSM-Pro method we proposed achieved outstanding results with multiple datasets, effectively demonstrating its efficacy. Through ablation studies, we further showcased the effectiveness and indispensability of each module within SBSM-Pro.

We defined a standard process termed PSD, which establishes the link between the physicochemical properties of amino acids and dictionaries for amino acid grouping. During the process of protein sequence alignment, the extensive size of the amino acid alphabet results in an excessive insertion of gaps. This compromises the alignment of sequences and subsequently obscures the accurate representation of similarities between protein sequences. PSD, by establishing a link between the original protein sequence and its structure, offers an effective solution to this challenge. The efficacy of this method has been validated through our experiments. This standardized process can be further utilized and developed by more researchers.

The SBSM-Pro method, which utilizes the LS distance and SW scoring to compute the similarities between proteins, is integrated with an SVM. By extracting multifaceted information from the raw protein sequences, it retains more information compared to traditional feature extraction techniques and achieves a higher accuracy rate.

We have proposed a novel MKL method, HCKDM-MKL, in an innovative manner into sequence classification. This method holds significant potential and promise. Numerous methods in biology are available for calculating protein similarity and generating corresponding kernel matrices. The introduction of MKL offers the possibility for SVMs to integrate information from different perspectives, such as ensemble learning, which will motivate more researchers to effectively improve the accuracy of sequence classification by designing a variety of sequence similarity metrics.

SBSM-Pro, which constructs sequence kernels from original sequences, has achieved outstanding results. However, numerous avenues for further exploration and in-depth research remain open. First, as part of our future endeavours, we plan to develop a graphical user interface to promote our software and make it more convenient for more biologists to use. Furthermore, we propose to analyse the structural and functional attributes of proteins, thereby assessing the similarity between protein sequences from these two perspectives. We believe that the construction of structural and functional kernels, coupled with the MKL method we proposed, has the potential to further enhance the performance of SBSM-Pro. In addition, it is important to note that SBSM-Pro was originally designed for bio-sequences. Apart from protein sequences, it should also encompass the classification tasks of DNA and RNA sequences.

In summary, SBSM-Pro, a classification model designed specifically for biological sequences, has achieved outstanding results. This pioneering work, characterized by its scalability, will inspire an increasing number of researchers to delve into related studies. These researchers can explore methods for measuring the similarity between protein sequences from various perspectives, generate similarity kernels, and integrate them into models through MKL methods. Additionally, they can utilize the existing models to assist or even guide biological experiments, probing into the potential information of biological sequences. SBSM-Pro is available for access at http://lab.malab.cn/soft/SBSM-Pro/.

\section{Materials and Methods}\label{Materials and Methods}

\begin{figure*}[!ht]
\centering
\includegraphics[width=1\textwidth]{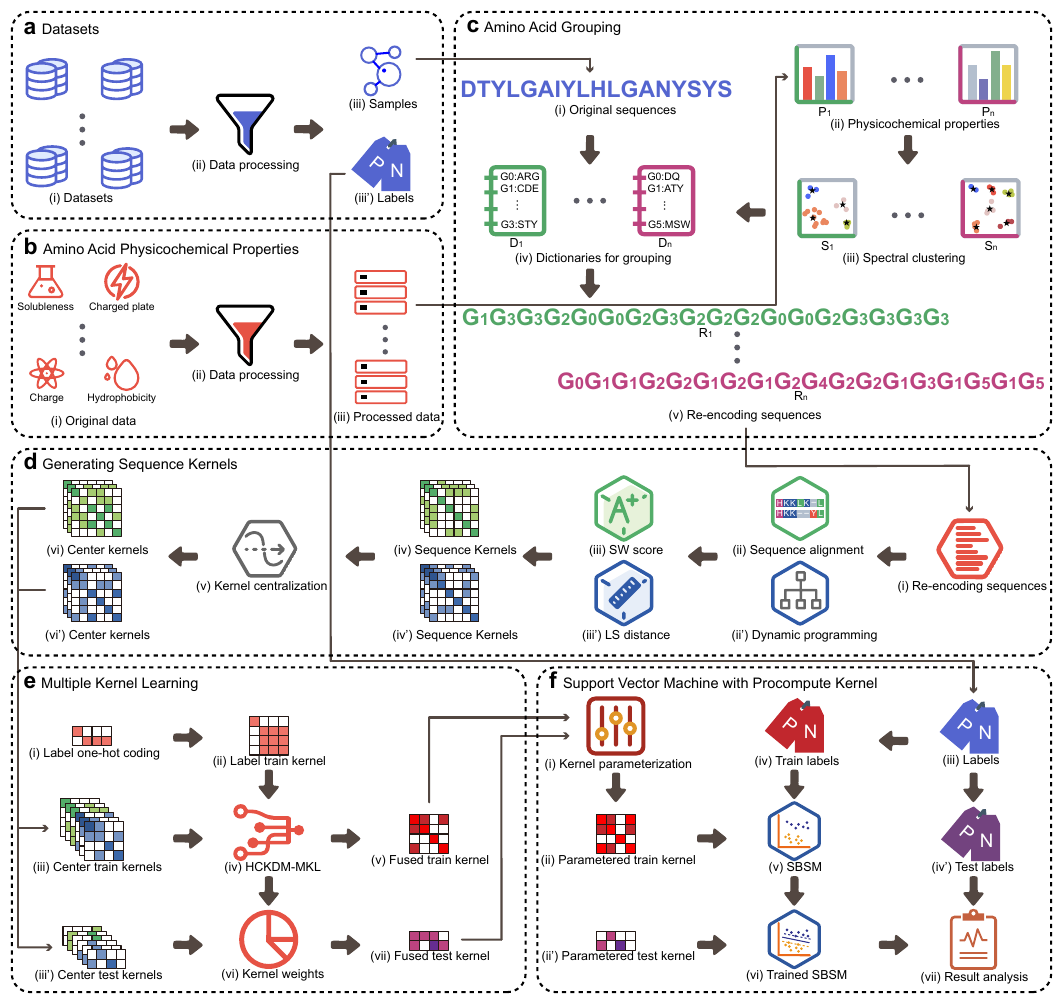}
\caption{\textbf{Overview of SBSM-Pro.} \textbf{a}, Datasets. We collected 10 commonly used datasets, and through data processing, we obtained 10 sets of corresponding labels and samples. \textbf{b}, Amino acid physicochemical properties. We collected physicochemical property data and subsequently processed these data, resulting in the processed data. \textbf{c}, Amino acid grouping. The physicochemical properties of amino acids obtained from step (b) were processed through spectral clustering to yield dictionaries for grouping. This standard process is defined as PSD. The original protein sequence samples derived from step (a) were transformed into a re-encoding sequence through the PSD. \textbf{d}, Sequence kernel generation. The re-encoded sequence obtained in step (c) underwent processing using sequence alignment and dynamic programming to calculate the SW scores and LS distances, respectively. This resulted in two sets of sequence kernels. After a kernel centralization process, the original sequence kernels from both sets were transformed into their corresponding central kernels. \textbf{e}, Multiple kernel learning. The label train kernel and the center train kernel, obtained from step (d), were input into the HCKDM-MKL to obtain kernel weights, which were then used to generate the fused train kernel. The kernel weights were then directly utilized to fuse the center test kernels, resulting in the fused test kernel. \textbf{f}, SVM with precompute kernel. The training labels and the parameterized training kernel were used to train the SVM. The parameterized test kernel was then input into the trained SVM to produce predicted results. Finally, these results were combined with the test labels for result analysis.}\label{SBSM-Pro}
\end{figure*}

In this section, we will delve into the methodologies associated with SBSM-Pro. Figure \ref{SBSM-Pro} provides an overview of SBSM-Pro. The first step involves collecting relevant datasets for protein identification. After collecting these datasets, they underwent processing, resulting in the creation of multiple sets of protein samples with their respective labels. Subsequently, we retrieved physicochemical property data of amino acids from the available literature. These data were also preprocessed. Finally, the amino acid physicochemical properties were subjected to spectral clustering, giving rise to the dictionaries for grouping. The original protein sequences were then transformed into re-encoding sequences by amino acid grouping in accordance with the corresponding dictionary. To gauge the similarity among these reencoding sequences, we employed sequence alignment techniques along with dynamic programming methods. Central kernels were derived by applying suitable kernel processing techniques. We proposed an innovative MKL strategy to fuse these central kernels. The fused central kernel was subsequently fed into SBSM-Pro for classification, ultimately leading to the final classification outcome.

\subsection{Datasets}

To evaluate SBSM-Pro, we collected 10 different protein classification datasets, including the identification of protein functions and posttranslational modifications (PTMs). The datasets encompass various protein functionalities, such as DNA-binding proteins (DBPs), type III secreted effectors (T3SEs), and phage virion proteins (PVPs), contributing to our understanding of genetic encoding, host‒pathogen interactions, and virus‒host relationships, respectively. Regarding posttranslational modifications (PTMs), we considered protein tyrosine sulfation sites (PTSS), protein s-nitrosylation sites (PSNS), protein lysine glutarylation sites (PLGS), and protein carbonylation sites (PCS). These PTMs play significant roles in modifying the behavioural properties of proteins and are implicated in numerous cellular processes, including metabolic regulation, redox reactions, and biological processes linked with various diseases. These datasets allow for a comprehensive evaluation of our SBSM-Pro, providing robust validation of our model through the identification of protein function and PTMs.

\begin{figure}[!ht]%
\centering
\includegraphics[width=1\textwidth]{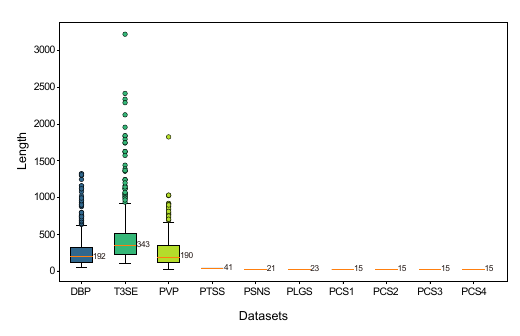}
\caption{\textbf{Boxplots of protein sequence lengths for different datasets.} The three boxplots on the left illustrate the distribution of protein sequence lengths in protein function identification.  In contrast, the seven boxplots on the right have been rendered as a single line, representing the fact that in the protein PTM identification datasets, the length of all proteins within each dataset share identical lengths.}\label{boxplots}
\end{figure}

We collected a set of commonly used datasets for protein classification, including 3 for protein function identification and 7 for PTM identification. We then analysed the protein sequence lengths and presented them as a box plot, as shown in Figure \ref{boxplots}. It is worth noting that the protein sequence lengths in each PTM identification dataset are consistent. However, the lengths of sequences for protein function identification vary.

In assessing two machine learning algorithms, it is crucial to employ the same training and testing sets for evaluation. This methodology eliminates variations that may arise from different data partitions, ensuring a reliable and fair comparison between algorithms. In our study, we categorized the collected datasets into two types: Type I and Type II. With regards to Type I datasets, it has been a customary practice in previous research to assess models using pre-partitioned datasets. We adhered to this practice to guarantee fairness in the comparison of the algorithms. Type II datasets, in contrast, do not have a predefined division into training and testing sets. For these datasets, we utilized the method of 10-fold cross-validation for model evaluation, in line with approaches utilized in prior studies.

The fundamental step of cross-validation involves partitioning the entire dataset into $K$ subsets. In each iteration, one subset is designated as the testing set, while the remaining $K-1$ subsets serve as the training set, yielding model evaluation outcomes. This process is executed $K$ times, ensuring a different testing set for each run. Consequently, $K$ model evaluation outcomes are obtained, and the final model performance assessment is derived from their average. A summary of the Type I and Type II datasets is shown in Tables \ref{datasets of type I} and \ref{datasets of type II}, respectively. 

\begin{table}[!ht]
\caption{Summary of datasets of type I utilized in the research.}\label{datasets of type I}
\begin{tabular}{llcccc}
\toprule
&& \multicolumn{2}{@{}c@{}}{Training set} & \multicolumn{2}{@{}c@{}}{Testing set} \\\cmidrule{3-4}\cmidrule{5-6}
Dataset & Description & Positive & Negative & Positive & Negative \\
\midrule
DBP & DNA-binding proteins\footnotemark[1] & 525 & 550 & 93 & 93\\
T3SE & Type III secreted effectors\footnotemark[2] & 309 & 310 & 42 & 34\\
PVP & Phage virion proteins\footnotemark[3] & 99 & 208 & 30 & 64\\
PTSS & Protein tyrosine sulfation sites\footnotemark[4] & 200 & 420 & 80 & 80\\
PSNS & Protein S-nitrosylation sites\footnotemark[5] & 731 & 810 & 43 & 121\\
PLGS & Protein lysine glutarylation sites\footnotemark[6] & 400 & 1703 & 44 & 203\\
\botrule
\end{tabular}
\footnotetext[1]{The dataset is obtained from the study\cite{Dataset1}.}
\footnotetext[2]{The dataset is obtained from the study\cite{Dataset2}.}
\footnotetext[3]{The dataset is obtained from the study\cite{Dataset3}.}
\footnotetext[4]{The dataset is obtained from the study\cite{Dataset4}.}
\footnotetext[5]{The dataset is obtained from the study\cite{Dataset5}.}
\footnotetext[6]{The dataset is obtained from the study\cite{Dataset6}.}
\end{table}

\begin{table}[!ht]
\caption{Summary of datasets of type II utilized in the research.}\label{datasets of type II}%
\begin{tabular}{llcc}
\toprule
Index & Description  & Positive & Negative\\
\midrule
PCS1 & Protein carbonylation sites 1\footnotemark[1] & 300 & 1949\\
PCS2 & Protein carbonylation sites 2\footnotemark[1] & 126 & 792\\
PCS3 & Protein carbonylation sites 3\footnotemark[1] & 136 & 847\\
PCS4 & Protein carbonylation sites 4\footnotemark[1] & 121 & 732\\
\botrule
\end{tabular}
\footnotetext[1]{The dataset is obtained from the study\cite{Dataset7to10}.}
\end{table}

\subsection{Physicochemical properties of amino acids}\label{Physicochemical properties of amino acids}

Proteins are composed of amino acids, fundamental organic compounds in biological processes. Each amino acid molecule consists of an amino group, a carboxyl group, a hydrogen atom, and a side chain. This particular structure of amino acids gives rise to various physicochemical properties.

We collated data on the physicochemical properties of amino acids from previous studies, which are frequently employed in bioinformatics research, including alpha-carbon positions (ACP) \cite{PhysicochemicalProperties1}, hydrophobicity (H) \cite{PhysicochemicalProperties2, PhysicochemicalProperties10}, secondary structure (SS) \cite{PhysicochemicalProperties3, PhysicochemicalProperties5}, non-bonded energy (NBE) \cite{PhysicochemicalProperties4}, membrane regions (MR) \cite{PhysicochemicalProperties6}, polarity and bulkiness (PB) \cite{PhysicochemicalProperties7}, chemical structure (CS) \cite{PhysicochemicalProperties8}, mean polarities (MP) 
 \cite{PhysicochemicalProperties9}, and side-chain (SC) \cite{PhysicochemicalProperties11, PhysicochemicalProperties12}.

These properties were numerically represented and retained for the purpose of generating dictionaries for grouping via spectral clustering. Regrettably, the data are not fully complete, necessitating further processing to ensure their usability and integrity.

Biological factors can often result in unusable or incomplete data. For example, the simplicity of the side chains in alanine and glycine, composed of a methyl group and a hydrogen atom, respectively, may result in a less pronounced impact during detailed side chain analysis compared to more complex amino acids. This often results in missing data, manifesting as not applicable (NA) in the numerical values for the physicochemical properties of these amino acids. Directly assigning a specific value, such as zero, to missing data could result in a loss of accuracy and interpretability. Thus, we opted to eliminate data entries for amino acids' physicochemical properties containing "NA". The processed data is shown in Table \ref{Processed data}. For reference, the removed data entries can be found in Appendix \ref{Supplementary material} Tabel \ref{Removed data}.

\begin{table}[!ht]
\caption{Summary of processed physicochemical properties of amino acids.}\label{Processed data}%
\begin{tabular}{lcccccccccc}
\toprule
& ACP\footnotemark[1]  & H1\footnotemark[2] & SS1\footnotemark[3] & NBE\footnotemark[4] & SS2\footnotemark[5] & MR\footnotemark[6] & PB\footnotemark[7] & CS\footnotemark[8] & MP\footnotemark[9] & H2\footnotemark[10]\\
Amino acid & $P_1$ & $P_2$ & $P_3$ & $P_4$ & $P_5$ & $P_6$ & $P_7$ & $P_8$ & $P_9$ & $P_{10}$ \\
\midrule
Ala (A) & 1.6 & 87 & 0.8 & -0.491 & 16 & 9.36 & 9.9 & 0.33 & -0.06 & -0.26 \\
Arg (R) & 0.9 & 81 & 0.96 & -0.554 & -70 & 0.27 & 4.6 & -0.176 & -0.84 & 0.08 \\
Asn (N) & 0.7 & 70 & 1.1 & -0.382 & -74 & 2.31 & 5.4 & -0.233 & -0.48 & -0.46 \\
Asp (D) & 2.6 & 71 & 1.6 & -0.356 & -78 & 0.94 & 2.8 & -0.371 & -0.8 & -1.3 \\
Cys (C) & 1.2 & 104 & 0 & -0.67 & 168 & 2.56 & 2.8 & 0.074 & 1.36 & 0.83 \\
Gln (Q) & 0.8 & 66 & 1.6 & -0.405 & -73 & 1.14 & 9 & -0.254 & -0.73 & -0.83 \\
Glu (E) & 2 & 72 & 0.4 & -0.371 & -106 & 0.94 & 3.2 & -0.409 & -0.77 & -0.73 \\
Gly (G) & 0.9 & 90 & 2 & -0.534 & -13 & 6.17 & 5.6 & 0.37 & -0.41 & -0.4 \\
His (H) & 0.7 & 90 & 0.96 & -0.54 & 50 & 0.47 & 8.2 & -0.078 & 0.49 & -0.18 \\
Ile (I) & 0.7 & 105 & 0.85 & -0.762 & 151 & 13.73 & 17.1 & 0.149 & 1.31 & 1.1 \\
Leu (L) & 0.3 & 104 & 0.8 & -0.65 & 145 & 16.64 & 17.6 & 0.129 & 1.21 & 1.52 \\
Lys (K) & 1 & 65 & 0.94 & -0.3 & -141 & 0.58 & 3.5 & -0.075 & -1.18 & -1.01 \\
Met (M) & 1 & 100 & 0.39 & -0.659 & 124 & 3.93 & 14.9 & -0.092 & 1.27 & 1.09 \\
Phe (F) & 0.9 & 108 & 1.2 & -0.729 & 189 & 10.99 & 18.8 & -0.011 & 1.27 & 1.09 \\
Pro (P) & 0.5 & 78 & 2.1 & -0.463 & -20 & 1.96 & 14.8 & 0.37 & 0 & -0.62 \\
Ser (S) & 0.8 & 83 & 1.3 & -0.455 & -70 & 5.58 & 6.9 & 0.022 & -0.5 & -0.55 \\
Thr (T) & 0.7 & 83 & 0.6 & -0.515 & -38 & 4.68 & 9.5 & 0.136 & -0.27 & -0.71 \\
Trp (W) & 1.7 & 94 & 0 & -0.839 & 145 & 2.2 & 17.1 & -0.011 & 0.88 & -0.13 \\
Tyr (Y) & 0.4 & 83 & 1.8 & -0.656 & 53 & 3.13 & 15 & -0.138 & 0.33 & 0.69 \\
Val (V) & 0.6 & 94 & 0.8 & -0.728 & 123 & 12.43 & 14.3 & 0.245 & 1.09 & 1.15 \\
\botrule
\end{tabular}
\footnotetext[1]{The dataset is obtained from the study\cite{PhysicochemicalProperties1}.}
\footnotetext[2]{The dataset is obtained from the study\cite{PhysicochemicalProperties2}.}
\footnotetext[3]{The dataset is obtained from the study\cite{PhysicochemicalProperties3}.}
\footnotetext[4]{The dataset is obtained from the study\cite{PhysicochemicalProperties4}.}
\footnotetext[5]{The dataset is obtained from the study\cite{PhysicochemicalProperties5}.}
\footnotetext[6]{The dataset is obtained from the study\cite{PhysicochemicalProperties6}.}
\footnotetext[7]{The dataset is obtained from the study\cite{PhysicochemicalProperties7}.}
\footnotetext[8]{The dataset is obtained from the study\cite{PhysicochemicalProperties8}.}
\footnotetext[9]{The dataset is obtained from the study\cite{PhysicochemicalProperties9}.}
\footnotetext[10]{The dataset is obtained from the study\cite{PhysicochemicalProperties10}.}
\end{table}

\subsection{Amino acid grouping}

In this section, we introduce the approach of grouping amino acids by using their physicochemical properties. First, we define a set of physicochemical properties that capture the essential characteristics of the protein sequences. These properties serve as the basis for subsequent analyses. Next, spectral clustering techniques are applied to partition the protein sequences based on their physicochemical similarities. This step helps to identify groups or clusters of proteins that share similar properties. Finally, we construct dictionaries to represent each protein group, capturing the underlying patterns and relationships within the clusters. By leveraging this process, we are able to effectively encode and represent protein sequences in a more meaningful and compact manner, enabling enhanced analysis and interpretation of protein data.

\subsubsection{Spectral clustering}

Spectral clustering, a graph theory-based clustering methodology, utilizes spectral information (i.e., eigenvectors) for data segmentation. Renowned for its robustness and adaptability, this data partitioning technique has garnered widespread attention in recent years.

Consider a set comprising $p$ distinct data points, which are clustered into $k_{c}$ clusters through spectral clustering. We first construct a similarity matrix $\bf{S}\in\bf{R}^{p\times p}$. The most prevalent similarity measure implemented is the Gaussian kernel of the Euclidean distance. Hence, the elements of matrix $\bf{S}$ can be computed using the following equation:

\begin{equation}
    S_{ij} = \exp\left(-\gamma \|x_i - x_j\|^2\right)
\end{equation}
where \(x_i\) and \(x_j\) are the data points. $\gamma$ is the coefficient of the kernel function, which effectively quantifies the decay rate of the similarity and determines how rapidly the similarity between data points diminishes as their distance increases.

Degree matrix $\bf{D}$ is defined as a diagonal matrix that satisfies $\bf{D}\in\bf{R}^{p\times p}$, and its elements can be calculated as

\begin{equation}
    \left[D\right]_{i,j}=\sum_{j=1}^{p}\left[S\right]_{i,j}
\end{equation}

Then, the graph Laplacian matrix $\bf{L}\in\bf{R}^{p\times p}$ is defined as

\begin{equation}
    \bf{L}=\bf{D}-\bf{S},
\end{equation}

Next, we proceed with the eigendecomposition of the Laplacian matrix. This decomposes the Laplacian matrix into a set of eigenvalues and their corresponding eigenvectors, thereby offering a more tractable framework for our subsequent analysis. Given that the Laplacian matrix is a real symmetric matrix, it is pertinent to note that all its eigenvalues are real numbers.

Subsequently, we select the $k_{c}$ smallest eigenvalues and form a matrix $\bf{U}\in\bf{R}^{p\times k_{c}}$ with corresponding eigenvectors as columns. The matrix $\bf{U}$ is row-normalized to obtain the matrix $\bf{T}\in\bf{R}^{p\times k_{c}}$. This involves scaling each row such that the sum of squares of all elements in a row equals one. Such normalization facilitates the clustering of data points within the same class while maximizing the distance between data points from different classes. Consequently, this enhances the effectiveness of the clustering process. We can conceptualize each row in the matrix $\bf{T}$ as an individual data point and then apply the K-means algorithm for clustering to derive the results. 

However, in our research, the task is to cluster 20 different amino acids. Unfortunately, we lack knowledge regarding the appropriate number of clusters to form, necessitating a metric to evaluate the effectiveness of our clustering efforts.

The Calinski‒Harabasz index (CHI), also known as the variance ratio criterion, is a commonly utilized metric for evaluating the outcomes of cluster analysis. It quantifies both the compactness within clusters and the separation between clusters. A higher value of the CHI suggests superior clustering performance.

Our dataset comprises $p$ elements, which have been clustered into $k_{c}$ clusters through spectral clustering. The evaluation metric $CHI$ for this particular clustering outcome can be calculated utilizing the following equation:

\begin{equation}
    CHI = \frac{{trace\left( {{B_k}} \right)}}{{trace\left( {{W_k}} \right)}} \times \frac{{{p} - k_{c}}}{{k_{c} - 1}}
\end{equation}
where $B_k$ and $W_k$ are the between-group dispersion matrix and within-cluster dispersion matrix, respectively. 

We define the center of $E$ as $C_E$. For a particular cluster $q$, its center is represented as $c_q$. The set of all data points contained within cluster $q$ is defined as $C_q$, with $n_q$ representing the number of elements in the set $C_q$. Subsequently, $B_k$ and $W_k$ can be calculated using the following equations:

\begin{equation}
    {B_k} = \sum\limits_{q = 1}^k {{n_q}\left( {{c_q} - {c_E}} \right)} {\left( {{c_q} - {c_E}} \right)^T}
\end{equation}

\begin{equation}
    {W_k} = \sum\limits_{q = 1}^k {\sum\limits_{x \in {C_q}} {\left( {x - {c_q}} \right)} } {\left( {x - {c_q}} \right)^T}
\end{equation}

\subsubsection{Dictionaries for grouping}

For a specific amino acid with distinct physicochemical properties, we employ spectral clustering to perform clustering analysis. The clustering process results in the formation of $k$ clusters, each containing at least one amino acid. We consider each cluster as a distinct group, defined as a set. An illustrative example is provided below:
\begin{equation}\label{G1}
    {G_1} = \left\{ {A,W} \right\}
\end{equation}

Due to the existence of $k$ clusters, there are $k$ sets, similar to those in Equation \ref{G1}. These sets satisfy the following equation:
\begin{equation}
    AA = \bigcup_{i=1}^{k} G_i
\end{equation}
where AA represents a set of 20 amino acids.

Next, we consider $k$ groups as $k$ dictionary entries and include them in a set, resulting in the formation of dictionaries for grouping that is defined as follows:

\begin{equation}
    {D_1} = \left\{ {{G_1}, \cdots ,{G_K}} \right\}
\end{equation}

In accordance with a specific dictionary, the amino acid residues of the original protein sequence are replaced by the group number in which they are located, resulting in the re-encoding protein sequence.

The physicochemical properties of amino acids are obtained after data processing, and corresponding clustering results are obtained through spectral clustering. According to the clustering results, dictionaries for grouping can be generated. There is a one-to-one correspondence between them. We define this standard process as PSD. The PSD process reduces the alphabet of the original protein sequence. In biology, amino acid residues within proteins may undergo substitutions. However, some of these substitutions may have minimal impact on the protein’s function or even no effect at all. By grouping amino acids using PSD, we can reduce such noise, facilitating subsequent sequence alignment. Additionally, this approach can link the original sequence of the protein to its structure, enabling a more accurate representation of interprotein distances.

\subsection{Generating Sequence Kernels}

In the previous section, we re-encoded the protein sequences. In this section, we generate the SW score and LS distance through sequence alignment and dynamic programming, respectively. Subsequently, a series of transformations, such as normalization and centralization, are applied to produce the sequence similarity kernel.

\subsubsection{Smith-Waterman alignment}

The SW algorithm is a widely used sequence alignment method in bioinformatics for identifying optimal local alignments between two sequences. Using this method, the similarity between proteins can be calculated.

To perform sequence alignment between two protein sequences, denoted as $Sx_i$ and $Sy_i$, and compute their Smith-Waterman (SW) scores, we employ the SW algorithm. The core of this algorithm can be formulated using a scoring matrix $A$, where each element $A_{i,j}$ represents the best score for aligning the prefixes of the two sequences up to positions $i$ and $j$. The equation for calculating the elements of the scoring matrix $A$ is as follows:

\begin{equation}
    {\left[ A \right]_{i,j}} = \max \left\{ \begin{array}{l}
{\left[ A \right]_{i,j - 1}} - g, \: if \: j>0 \: and \: i \geq 0\\
{\left[ A \right]_{i - 1,j}} - g, \: if \: j \geq 0 \: and \: i > 0\\
{\left[ A \right]_{i - 1,j - 1}} + p\left( {i,j} \right), \: if \: j > 0 \: and \: i > 0\\
\end{array} \right.
\end{equation}
where $ A_{i,j-1} $ denotes a gap at position $ j $ of sequence $ Sy $, $ A_{i-1,j} $ denotes a gap at position $ i $ of sequence $ Sx $, and $ A_{i-1,j-1} $ indicates an alignment without gaps at positions $ i $ and $ j $. The $ p_{ij} $ is a function that allocates scores based on matches or mismatches at positions $ i $ and $ j $. It is defined as follows:

\begin{equation}
    p\left( {i,j} \right) = \left\{ \begin{array}{l}
{m_1}, \: if \: Sx_i = Sy_j\\
{m_2}, \: if \: Sx_i \neq Sy_j
\end{array} \right.
\end{equation}
where ${m_1}$ and ${m_2}$ represent the scores for a match and a mismatch between elements at positions $i$ and $j$, respectively.

After computing the scoring matrix, a traceback can be performed. In contrast to the Needleman‒Wunsch algorithm used for global alignment, which backtracks from the bottom right corner of the scoring matrix to the bottom left corner, the SW algorithm initiates the traceback from the highest value within the scoring matrix and stops when it reaches a score of zero, thereby identifying the optimal local alignment. However, our primary goal in incorporating the SW algorithm is to obtain the SW score. Therefore, we do not need to perform the traceback process. Instead, we simply choose the maximum value from the scoring matrix as the SW score.

\begin{figure}[!ht]%
\centering
\includegraphics[width=0.7\textwidth]{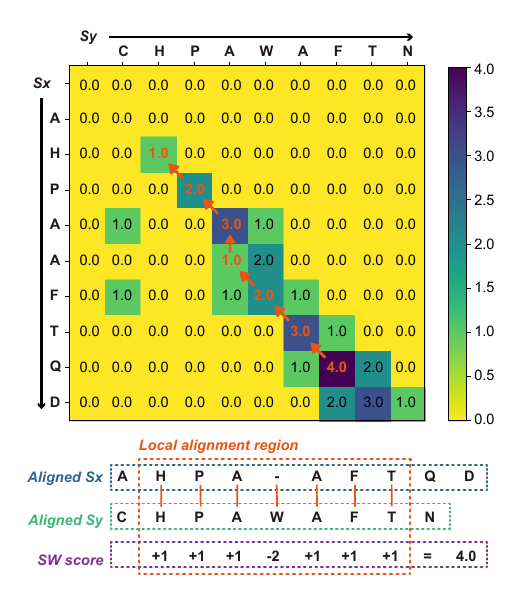}
\caption{\textbf{Schematic of the SW algorithm.} The scoring matrix, generated using the Smith-Waterman (SW) algorithm for the provided sequence, is displayed above, with the scoring matrix values color-mapped. The path of backtracking after dynamic programming is indicated by red numbers and red arrows. The following presents two protein sequences, $S_x$ and $S_y$, illustrated in blue and green dashed boxes, respectively. The orange dashed box signifies the local alignment region between these two sequences. The SW score is calculated within the local alignment region, denoted by the purple dashed box.}\label{SW schematic}
\end{figure}

The schematic diagram of the SW algorithm is shown in Figure \ref{SW schematic}. In the example diagram, a gap is introduced at the fifth position of protein sequence $Sx$. Starting from the second amino acid of both proteins, a local alignment region comprising seven amino acids emerges, resulting in a final SW score of 4.

The following equation presents the normalized SW score for the protein sequences $S_x$ and $S_y$:

\begin{equation}
    {SW^ * }\left( {{S_x},{S_y}} \right) = \frac{{SW\left( {{S_x},{S_y}} \right)}}{{\max \left( {{l_x},{l_y}} \right)}}
\end{equation}
where ${SW\left( {{S_x},{S_y}} \right)}$ and $ {SW^ * }\left( {{S_x},{S_y}} \right)$ represent the original and normalized SW scores, respectively. Meanwhile, $l_x$ and $l_y$ correspond to the lengths of protein sequences $S_x$ and $S_y$, respectively. 

By computing the SW scores between all pairs of sequences in the sample set, we store the results in the symmetric matrix $K_{SW}$, thus obtaining a protein similarity kernel based on the SW algorithm.

\subsubsection{Levenshtein distance}

The LS distance, also known as the edit distance, is a widely used metric for measuring the dissimilarity between two strings. It quantifies the minimum number of single-character operations required to transform one string into another. These operations include insertion, deletion, and substitution.

The LS algorithm, similar to the SW algorithm, is computed using dynamic programming. For two protein sequences $S_x$ and $S_y$ with lengths $l_x$ and $l_y$ respectively, a matrix  ${\bf{M}} \in {{\bf{R}}^{l_x \times l_y}}$ must first be generated when calculating their LS distance.

The LS algorithm, similar to the SW algorithm, is computed through dynamic programming. When calculating the LS distance between two protein sequences $S_x$ and $S_y$, each with lengths $l_x$ and $l_y$ respectively, a matrix ${\bf{M}} \in {{\bf{R}}^{l_x \times l_y}}$ is initially generated. The matrix serves as a dynamic programming table that stores the intermediate distances between substrings. Initialization of the matrix involves setting the first row and column to the respective indices, representing the cost of transforming an empty string into the corresponding prefix or vice versa. The subsequent entries in the matrix are filled iteratively by comparing characters at each position and determining the minimum cost of transforming the prefixes. The final value in the bottom right corner of the matrix represents the LS distance between the two input strings. The LS distances for all protein sequence pairs were calculated, resulting in the LS distance matrix, $\bf{D_{LS}}$.

Given that the LS distance measures the dissimilarity between two protein sequences, a protein similarity kernel based on the LS distance ${K_{LS}}$ can be obtained through normalization methods. The equation is as follows:

\begin{equation}
    \bf{K_{LS}} = 1 - \frac{{\bf{D_{LS}} - \min (\bf{D_{LS}})}}{{\max (\bf{D_{LS}}) - \min (\bf{D_{LS}})}}
\end{equation}
where $\max (\bf{D_{LS}})$ and $\min (\bf{D_{LS}})$ represent the maximum and minimum elements in the matrix $\bf{D_{LS}}$, respectively.

\subsection{Multiple kernel learning}

MKL methods have gained significant attention in the field of machine learning due to their ability to effectively model complex relationships in data. These methods extend the traditional single kernel approach by combining multiple kernels to capture information on different aspects of the data. By employing MKL methods, we are able to leverage complementary information from different feature representations, leading to improved predictive performance. In the MKL model, the fused kernel is derived by determining the kernel weights.

For a dataset comprising $N$ samples, we construct a set of $p$ kernel matrices by building the LS kernel and the SW kernel, as defined below:

\begin{equation}
    \bf{K} = \left\{ {{\bf{K}_1},{\bf{K}_2}, \cdots ,{\bf{K}_p}} \right\},{\bf{K}_p} \in {\bf{R}^{N \times N}}
\end{equation}

The objective of MKL is to determine the kernel weights, denoted as $\bf{\beta}$. Its definition is as follows:

\begin{equation}
    \bf{\beta}  = \left[ {{\beta _1},{\beta _2}, \cdots ,{\beta _p}} \right]
\end{equation}

For a set of N training samples, there are corresponding sample labels represented by $ \bf{L} \in R^{N \times 1} $. These labels are transformed into a one-hot encoding, denoted as $ \bf{Y}_{train} $. We define the target kernel as $ \bf{U} \in R^{N \times N}$ , with its equation given as follows:

\begin{equation}
    \bf{U} = \bf{Y}_{train} \bf{Y}_{train}^T 
\end{equation}

\begin{figure}[!ht]%
\centering
\includegraphics[width=1\textwidth]{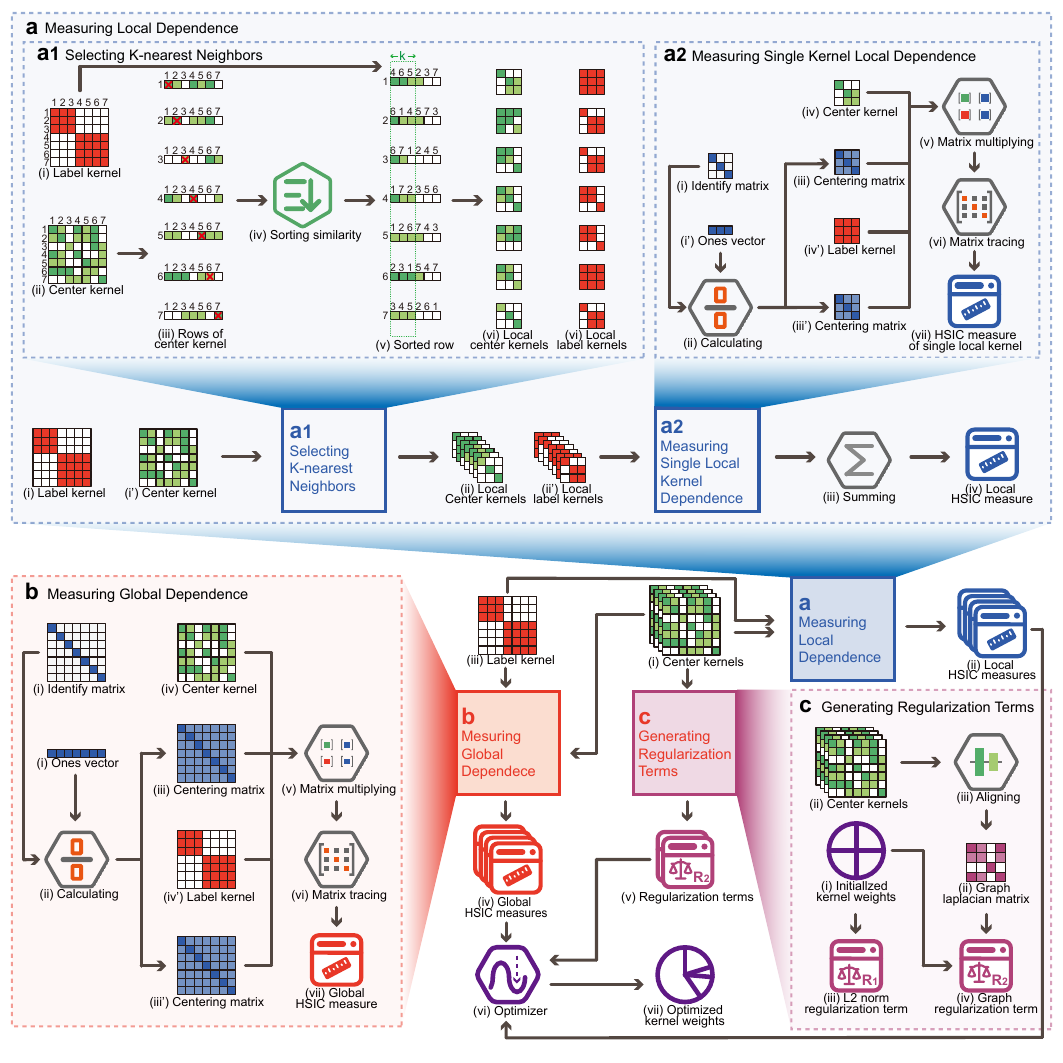}
\caption{\textbf{Overview of HCKDM-MKL.} \textbf{a}, Measuring local dependence. By sorting the similarity of each row in the N×N center kernel matrix, we can obtain N groups, each containing K-nearest neighbors. By extracting the rows and columns of the center kernel and label kernel matrices according to N groups of K-nearest neighbors, we can obtain N local center kernels and local label kernels, respectively. The HSIC measure of a single local kernel is obtained by performing specific multiplication operations involving the center kernel, label kernel, and centering matrix, followed by computing the trace. The local HSIC measure is obtained by summing N outcomes. \textbf{b}, Measuring global dependence. After carrying out the specified multiplication operations involving the central kernel, the label kernel, and the centering matrix, the trace is calculated to derive the global HSIC measures. \textbf{c}, Generating regularization terms. The graph Laplacian matrix is obtained through the aligning center kernels. The initialized kernel weights directly yield the L2 norm regularization term, which, when combined with the graph Laplacian matrix, results in the graph regularization term. The obtained local HSIC measures, global HSIC measures, and regularization terms are input into the optimizer to yield the optimized kernel weights.}\label{HCKDM_MKL}
\end{figure}

\subsubsection{Hilbert‒Schmidt Independence Criterion}

In the domain of machine learning and statistics, assessing the independence between random variables is of paramount significance. The HSIC provides an efficient, nonparametric method for independence testing \cite{hsic}. This criterion assesses the dependency between two sets of variables by evaluating the discrepancy between joint and product distributions in a reproducing kernel Hilbert space (RKHS). Essentially, it quantifies independence by calculating the norm of the cross-covariance operator in the RKHS. Notably, the HSIC method performs well in high-dimensional spaces and with complex relationships, and it is capable of detecting nonlinear dependencies. As a criterion based on the kernel mean embeddings of distributions, HSIC is particularly useful in scenarios where the explicit form of the joint distribution is unknown or challenging to ascertain, and it has seen wide applications in areas such as feature selection, causality testing, and variable independence verification.

We define ${\bf{X}} = {\left\{ {{{\bf{x}}_1},{{\bf{x}}_2}, \cdots ,{{\bf{x}}_N}} \right\}^T} \in {{\bf{R}}^{N \times d}}$ as the original feature of $d$ dimensions of samples, and ${\bf{Y}} \in {{\bf{R}}^{N \times 1}}$ is the label of these samples. We can derive a series of observations from the probability distribution ${\Pr _{xy}}$, defined as
\begin{equation}
    Z \equiv \left\{ {\left( {{{\bf{x}}_1},{y_1}} \right),\left( {{{\bf{x}}_2},{y_2}} \right), \cdots ,\left( {{{\bf{x}}_N},{y_N}} \right)} \right\} \subseteq {\bf{X}} \times {\bf{Y}}
\end{equation}

HSIC calculates the cross-covariance operator on the domain ${\bf{X}} \times {\bf{Y}}$ to determine the independence between ${\bf{X}}$ and ${\bf{Y}}$. The feature set ${\bf{X}}$ and label set ${\bf{Y}}$ can be mapped to ${\bf{F}}$ and ${\bf{G}}$ by the mapping $\phi :{\bf{X}} \to {\bf{F}}$ and $\psi :{\bf{Y}} \to {\bf{G}}$. Then, we defined their expectations as ${\mu _{\bf{x}}}$ and ${\mu _y}$, respectively. The kernel function is as follows:
\begin{equation}
    k\left( {{{\bf{x}}_i},{{\bf{x}}_j}} \right) = \left\langle {\phi \left( {{{\bf{x}}_i}} \right),\phi \left( {{{\bf{x}}_j}} \right)} \right\rangle 
\end{equation}

Similarly, the kernel function of ${\bf{Y}}$ is defined as
\begin{equation}
    l\left( {{y_i},{y_j}} \right) = \left\langle {\psi \left( {{y_i}} \right),\psi \left( {{y_j}} \right)} \right\rangle 
\end{equation}

The following equation can be used to determine the cross-covariance operator ${C_{{\bf{x}}y}}$:
\begin{equation}
    {C_{{\bf{x}}y}} = {E_{{\bf{x}},y}}\left[ {\phi \left( {\bf{x}} \right) \otimes \psi \left( y \right)} \right] - {\mu _{\bf{x}}}{\mu _y}
\end{equation}
where ${E_{{\bf{x}},y}}$ denotes the common expectation of ${\bf{x}}$ and $y$. Then, we can write the HSIC operator is:
\begin{equation}
    HSIC\left( {{\bf{F}},{\bf{G}},{{\Pr }_{xy}}} \right) = \left\| {{C_{{\bf{x}}y}}} \right\|_{HS}^2
\end{equation}

Then, we define as the ${\bf{I}}$ identity matrix, and it satisfies ${\bf{I}} \in {{\bf{R}}^{N \times N}}$. By defining ${\bf{e}} = {\left[ {1,1, \cdots ,1} \right]^T} \in {{\bf{R}}^{1 \times N}}$, we can obtain
\begin{equation}
    {\bf{H}} \equiv {\bf{I}} - \frac{{{\bf{e}}{{\bf{e}}^T}}}{N}
\end{equation}

Note that ${\bf{H}}$ is the centering matrix, and it satisfies ${\bf{H}} \in {{\bf{R}}^{N \times N}}$. Then, we can make an empirical estimate of ${\bf{Z}}$ set as
\begin{equation}
\begin{array}{c}
    HSIC\left( {{\bf{F}},{\bf{G}},{\bf{Z}}} \right) = \frac{1}{{{N^2}}}tr({\bf{KU}}) - \frac{2}{{{N^3}}}{{\bf{e}}^T}{\bf{KUe}} + \frac{1}{{{N^4}}}{{\bf{e}}^T}{\bf{Ke}}{{\bf{e}}^T}{\bf{Ue}}\\
     = \frac{1}{{{N^2}}}\left[ {tr\left( {{\bf{KU}}} \right) - \frac{1}{N}tr\left( {{\bf{KUe}}{{\bf{e}}^T}} \right) - \frac{1}{N}tr\left( {{\bf{UKe}}{{\bf{e}}^T}} \right) + \frac{1}{{{N^2}}}tr({\bf{Ue}}{{\bf{e}}^T}{\bf{Ke}}{{\bf{e}}^T})} \right]\\
     = \frac{1}{{{N^2}}}tr\left[ {{\bf{K}}\left( {{\bf{I}} - \frac{1}{N}{\bf{e}}{{\bf{e}}^T}} \right){\bf{U}}\left( {{\bf{I}} - \frac{1}{N}{\bf{e}}{{\bf{e}}^T}} \right)} \right]\\
     = \frac{1}{{{N^2}}}tr({\bf{KHUH}}) \buildrel \Delta \over = HSIC({\bf{K}},{\bf{U}})
\end{array}
\end{equation}
where ${\bf{K}},{\bf{U}}  \in {{\bf{R}}^{N \times N}}$ are kernel matrices, $k\left( {{{\bf{x}}_i},{{\bf{x}}_j}} \right)$ and $l\left( {{y_i},{y_j}} \right)$. It's important to note that the value of $HSIC({\bf{K}},{\bf{U}})$ is associated with the dependence between ${\bf{K}}$ and ${\bf{U}}$, and a higher value indicates a stronger dependence between the two. In addition, it should be in the range of 0 to 1. When it is equal to 0, we think that ${\bf{K}}$ and ${\bf{U}}$ are independent or irrelevant.

\subsubsection{HCKDM-MKL}

First, we centralize the acquired kernel matrix to normalize the data, ensuring that the similarity or distance of each data point consistently influences the results. Moreover, by subtracting the mean values of rows and columns, the centered kernel emphasizes the similarity information. All these steps enhance the effectiveness of MKL algorithms. The equation is as follows:

\begin{equation}
    {\left( {{{\widehat {\bf{K}}}_p}} \right)_{i,j}} = {\left( {{{\bf{K}}_p}} \right)_{i,j}} - \frac{1}{N}\sum\limits_{i = 1}^N {{{\left( {{{\bf{K}}_p}} \right)}_{i,j}}}  - \frac{1}{N}\sum\limits_{j = 1}^N {{{\left( {{{\bf{K}}_p}} \right)}_{i,j}}}  + \frac{1}{{{N^2}}}\sum\limits_{i,j = 1}^N {{{\left( {{{\bf{K}}_p}} \right)}_{i,j}}}
\end{equation}
where $\bf{K}_p$ represents the original kernel of the $p^{th}$ similarity kernel and ${\widehat {\bf{K}}}_p$ denotes the centered kernel.

HCKDM leverages local kernels due to their computational efficiency in conducting convolution operations on the kernel matrix. A local kernel is a compact matrix employed to extract specific sample characteristics from the kernel matrix. Using local kernels allows us to achieve high precision in the kernel matrix dependency measure while minimizing the utilization of computational resources. The approach using local kernels has also been employed in other studies \cite{hkam}. Before using HSIC to measure all sample local kernels and label kernel, we first define ${{\bf{I}}_{local}} \in {{\bf{R}}^{k \times k}}$ and ${{\bf{e}}_{local}} = {\left[ {1,1, \cdots ,1} \right]^T} \in {{\bf{R}}^{1 \times k}}$. We can then calculate the quadratic matrix by the following equation: 
\begin{equation}
    {{\bf{H}}_{local}} \equiv {\bf{I}} - \frac{{{{\bf{e}}_{local}}{\bf{e}}_{local}^T}}{N} \in {{\bf{R}}^{k \times k}}
\end{equation}

Then, the dependence measures of local and label kernels for all samples $M_{local}$ can be calculated by HSIC, whose equation is shown below
\begin{equation}
\begin{array}{c}
    {M_{local}} = \frac{1}{N}\sum\limits_{i = 1}^N {HSIC\left( {{{\bf{\hat K}}^{*(i)}},{{\bf{U}}^{(i)}}} \right)} \\
     = \frac{1}{N}\sum\limits_{i = 1}^N {\frac{1}{{{k^2}}}} tr\left( {{{\bf{\hat K}}^{*(i)}}{{\bf{H}}_{local}}{{\bf{U}}^{(i)}}{{\bf{H}}_{local}}} \right)
\end{array}
\end{equation}

In contrast to local kernels, global kernels are employed to globally represent the characteristics of all kernel functions. To ensure compatibility in matrix dimensions during multiplication, it is necessary to initially define ${{\bf{I}}_{global}} \in {{\bf{R}}^{N \times N}} \in {{\bf{R}}^{N \times N}}$ and $ {{\bf{e}}_{global}} = {\left[ {1,1, \cdots ,1} \right]^T} \in {{\bf{R}}^{1 \times N}}$. Subsequently, we compute the centering matrix according to the equation presented below.
\begin{equation}
    {{\bf{H}}_{global}} \equiv {\bf{I}} - \frac{{{{\bf{e}}_{global}}{\bf{e}}_{global}^T}}{N} \in {{\bf{R}}^{N \times N}}
\end{equation}

\begin{equation}
\begin{array}{c}
    {M_{global}} = HSIC\left( {{{\bf{\hat K}}^*},{\bf{U}}} \right)\\
     = \frac{1}{{{N^2}}}tr\left( {{{\bf{\hat K}}^*}{{\bf{H}}_{global}}{\bf{U}}{{\bf{H}}_{global}}} \right)
\end{array}
\end{equation}

${R_1}$ is the $L_2$ norm regularization, and the equation is as follows: 
\begin{equation}
    {R_1} = {\nu _1}{\left\| {\bf{\beta }} \right\|^2}
\end{equation}

We define the Frobenius inner as
\begin{equation}
    {\left\langle {{\bf{K}},{\bf{U}}} \right\rangle _F} = tr\left( {{{\bf{K}}^T},{\bf{U}}} \right)
\end{equation}

Then, we defined ${\bf{W}}$ as the cosine similarity matrix between two kernels satisfying ${\bf{W}} \in {{\bf{R}}^{P \times P}}$ , and the equation is as follows:
\begin{equation}
    Aligned\left( {{\bf{K}},{\bf{U}}} \right) = \frac{{{{\left\langle {{\bf{K}},{\bf{U}}} \right\rangle }_F}}}{{{{\left\| {\bf{K}} \right\|}_F}{{\left\| {\bf{U}} \right\|}_F}}}
\end{equation}
where ${\left\| {\bf{K}} \right\|_F} = \sqrt {{{\left\langle {{\bf{K}},{\bf{K}}} \right\rangle }_F}} $ is the Frobenius norm. $\bf{{D_K}}$ is defined as a diagonal matrix that satisfies ${\bf{W}} \in {{\bf{R}}^{P \times P}}$, and its elements can be calculated as
\begin{equation}
    {\left[ {{D_K}} \right]_{i,j}} = \sum\limits_j^p {{{\left[ W \right]}_{i,j}}} 
\end{equation}

Then, the graph Laplacian matrix ${\bf{L}}_k$ is defined as
\begin{equation}
    {{\bf{L}}_k} = {{\bf{D}}_k} - {\bf{W}}
\end{equation}

We can write the Laplacian regular term as
\begin{equation}
\begin{array}{c}
    \sum\limits_{i,j}^p {{{\left( {{\beta _i} - {\beta _j}} \right)}^2}{W_{ij}}}  = \sum\limits_{i,j}^p {\left( {\beta _i^2 + \beta _j^2 - 2{\beta _i}{\beta _j}} \right){W_{ij}}} \\
     = \sum\limits_i^p {\beta _i^2{D_{ii}}}  + \sum\limits_j^p {\beta _j^2{D_{jj}}}  - 2\sum\limits_{i,j}^p {{\beta _i}{\beta _j}{W_{ij}}} \\
     = 2{{\bf{\beta }}^T}{{\bf{L}}_K}{\bf{\beta }}
\end{array}
\end{equation}

We define ${R_2}$ as the graph regularization term, which assists in smoothing the weights. The equation is as follows:
\begin{equation}
    {R_2} = {{\bf{\beta }}^T}{{\bf{L}}_K}{\bf{\beta }}
\end{equation}

We combine the regularization terms ${R_1}$ and ${R_2}$ to obtain the final regularization term ${R}$ as follows: 
\begin{equation}
\begin{array}{c}
    R = {R_1} + {R_2}\\
     = {\nu _1}{\left\| {\bf{\beta }} \right\|^2} + {\nu _2}{{\bf{\beta }}^T}{{\bf{L}}_K}{\bf{\beta }}
\end{array}
\end{equation}

Hence, we define a new kernel dependence measuring approach that concurrently considers global and local kernel dependence measurements and uses a parameter $\lambda \left( {0 \le \lambda  \le 1} \right)$ to establish a trade-off between these two types of kernel alignment. The hybrid dependence measuring between two kernel matrices is as follows:
\begin{equation}
    \max \lambda {M_{local}} + \left( {1 - \lambda } \right){M_{global}} - R
\end{equation}

Then, our method is transformed into an optimization problem, and the optimized fusion kernel can be obtained by maximizing the hybrid dependence, whose equation is shown as follows:
\begin{equation}
\begin{array}{c}
    \mathop {\max }\limits_{{\bf{\beta }},{{\bf{\hat K}}^*}} \lambda \frac{1}{N}\sum\limits_{i = 1}^N {\frac{1}{{{k^2}}}} tr\left( {{{\bf{{\hat K}}}^{*(i)}}{{\bf{H}}_{local}}{{\bf{U}}^{(i)}}{{\bf{H}}_{local}}} \right) + \left( {1 - \lambda } \right)\frac{1}{{{N^2}}}tr\left( {{{\bf{\hat K}}^*}{{\bf{H}}_{global}}{\bf{U}}{{\bf{H}}_{global}}} \right)\\
     - {\nu _1}{\left\| {\bf{\beta }} \right\|^2} - {\nu _2}{{\bf{\beta }}^T}{{\bf{L}}_K}{\bf{\beta }}\\
    st.{{\bf{\hat K}}^*} = \sum\limits_{i = 1}^p {{\beta _i}{{\bf{\hat K}}_i}} \\
    \sum\limits_{i = 1}^p {{\beta _i}}  = 1\\
    {\beta _i} \ge 0
\end{array}
\end{equation}

\subsection{Support vector machine with precomputed kernel}

The similarity kernel obtained from MKL first needs to be parameterized to be compatible with the SVM. Kernel parameterization offers the advantage of enhancing the performance of classifiers or regressors without increasing the computational complexity by mapping the data to a higher-dimensional feature space. Additionally, it can better handle nonlinear relationships among high-dimensional data and samples, thereby expanding the application scope of traditional linear methods. The kernel function of the $i$-th sequence is defined as.

\begin{equation}
    {\bf{K}}\left( {{\bf{x}},{{\bf{x}}_i}} \right) = \exp \left( {\alpha \frac{{s\left( {{\bf{x}},{{\bf{x}}_i}} \right) - {b_i}}}{{{b_i}}}} \right)
\end{equation}
where ${s\left( {{\bf{x}},{{\bf{x}}_i}} \right)}$ is a pairwise similarity measurement between ${\bf{x}}$ and ${{{\bf{x}}_i}}$ , ${{b_i}}$ is the maximum similarity measurement associated with the $i$-th support sequence and $\alpha$ is a constant. 

The given equation represents the dual form of the SVM optimization problem. 

\begin{equation}
\begin{array}{c}
\mathop {\max }\limits_\alpha  \sum\limits_{i = 1}^m {{\alpha _i}}  - \frac{1}{2}\sum\limits_{i = 1}^m {\sum\limits_{j = 1}^m {{\alpha _i}{\alpha _j}{y_i}{y_j}{{\hat K}^*}\left( {{{\bf{x}}_i},{{\bf{x}}_j}} \right)} } \\
s.t.\sum\limits_{i = 1}^m {{\alpha _i}} {y_i} = 0,\\
0 \le {\alpha _i} \le C,i = 1,2, \ldots ,m.
\end{array}
\end{equation}

This function is maximized with respect to $\alpha$, which is a vector of Lagrange multipliers. The first term of the objective function is the sum of all $\alpha_i$ from 1 to m. The second term is the dot product of the feature vectors $\bf{x}_i$ and $\bf{x}_j$ scaled by the corresponding $\alpha_i$, $\alpha_j$, and the labels $y_i$, $y_j$, summed over all pairs of data points. Through HCKDM-MKL, we obtain the fused kernel matrix ${\bf {\hat{K}}}^*$, which corresponds to the kernel function ${{\hat K}^*}\left( {{{\bf{x}}_i},{{\bf{x}}_j}} \right)$.

The first constraint ensures that the sum of $\alpha_i$ times the corresponding $y_i$ (the label of each data point) over all data points equals zero. The second constraint bounds each $\alpha_i$ to be nonnegative and no larger than a constant C for all data points. The constant C is a parameter for the SVM that controls the trade-off between maximizing the margin and minimizing the classification error.

\section*{Date availability}

In this paper, the datasets for the ten protein classifications are available on GitHub. The repository can be accessed at https://github.com/wyzbio/Support-Bio-sequence-Machine.

\section*{Code availability}

The source code of SBSM-Pro is freely available in the GitHub repository at https://github.com/wyzbio/Support-Bio-sequence-Machine.

\bibliography{sn-bibliography}% common bib file

\begin{thebibliography}{10}
\expandafter\ifx\csname url\endcsname\relax
  \def\url#1{\burl{#1}}\fi
\expandafter\ifx\csname urlprefix\endcsname\relax\def\urlprefix{URL }\fi
\providecommand{\bibinfo}[2]{#2}
\providecommand{\eprint}[2][]{\url{#2}}
\providecommand{\doi}[1]{\url{https://doi.org/#1}}
\bibcommenthead

\bibitem{PseKNC_General}
\bibinfo{author}{Chen, W.} \emph{et~al.}
\newblock \bibinfo{title}{{PseKNC-General: a cross-platform package for
  generating various modes of pseudo nucleotide compositions}}.
\newblock \emph{\bibinfo{journal}{Bioinformatics}}
  \textbf{\bibinfo{volume}{31}}, \bibinfo{pages}{119--120}
  (\bibinfo{year}{2014}).
\newblock \urlprefix\url{https://doi.org/10.1093/bioinformatics/btu602}.

\bibitem{PyFeat}
\bibinfo{author}{Muhammod, R.} \emph{et~al.}
\newblock \bibinfo{title}{{PyFeat: a Python-based effective feature generation
  tool for DNA, RNA and protein sequences}}.
\newblock \emph{\bibinfo{journal}{Bioinformatics}}
  \textbf{\bibinfo{volume}{35}}, \bibinfo{pages}{3831--3833}
  (\bibinfo{year}{2019}).
\newblock \urlprefix\url{https://doi.org/10.1093/bioinformatics/btz165}.

\bibitem{iFeature}
\bibinfo{author}{Chen, Z.} \emph{et~al.}
\newblock \bibinfo{title}{{iFeature: a Python package and web server for
  features extraction and selection from protein and peptide sequences}}.
\newblock \emph{\bibinfo{journal}{Bioinformatics}}
  \textbf{\bibinfo{volume}{34}}, \bibinfo{pages}{2499--2502}
  (\bibinfo{year}{2018}).
\newblock \urlprefix\url{https://doi.org/10.1093/bioinformatics/bty140}.

\bibitem{VisFeature}
\bibinfo{author}{Wang, J.} \emph{et~al.}
\newblock \bibinfo{title}{{VisFeature: a stand-alone program for visualizing
  and analyzing statistical features of biological sequences}}.
\newblock \emph{\bibinfo{journal}{Bioinformatics}}
  \textbf{\bibinfo{volume}{36}}, \bibinfo{pages}{1277--1278}
  (\bibinfo{year}{2019}).
\newblock \urlprefix\url{https://doi.org/10.1093/bioinformatics/btz689}.

\bibitem{POSSUM}
\bibinfo{author}{Wang, J.} \emph{et~al.}
\newblock \bibinfo{title}{{POSSUM: a bioinformatics toolkit for generating
  numerical sequence feature descriptors based on PSSM profiles}}.
\newblock \emph{\bibinfo{journal}{Bioinformatics}}
  \textbf{\bibinfo{volume}{33}}, \bibinfo{pages}{2756--2758}
  (\bibinfo{year}{2017}).
\newblock \urlprefix\url{https://doi.org/10.1093/bioinformatics/btx302}.

\bibitem{Rcpi}
\bibinfo{author}{Cao, D.-S.}, \bibinfo{author}{Xiao, N.}, \bibinfo{author}{Xu,
  Q.-S.} \& \bibinfo{author}{Chen, A.~F.}
\newblock \bibinfo{title}{{Rcpi: R/Bioconductor package to generate various
  descriptors of proteins, compounds and their interactions}}.
\newblock \emph{\bibinfo{journal}{Bioinformatics}}
  \textbf{\bibinfo{volume}{31}}, \bibinfo{pages}{279--281}
  (\bibinfo{year}{2014}).
\newblock \urlprefix\url{https://doi.org/10.1093/bioinformatics/btu624}.

\bibitem{protr}
\bibinfo{author}{Xiao, N.}, \bibinfo{author}{Cao, D.-S.}, \bibinfo{author}{Zhu,
  M.-F.} \& \bibinfo{author}{Xu, Q.-S.}
\newblock \bibinfo{title}{{protr/ProtrWeb: R package and web server for
  generating various numerical representation schemes of protein sequences}}.
\newblock \emph{\bibinfo{journal}{Bioinformatics}}
  \textbf{\bibinfo{volume}{31}}, \bibinfo{pages}{1857--1859}
  (\bibinfo{year}{2015}).
\newblock \urlprefix\url{https://doi.org/10.1093/bioinformatics/btv042}.

\bibitem{DiProDB}
\bibinfo{author}{Friedel, M.}, \bibinfo{author}{Nikolajewa, S.},
  \bibinfo{author}{Sühnel, J.} \& \bibinfo{author}{Wilhelm, T.}
\newblock \bibinfo{title}{{DiProDB: a database for dinucleotide properties}}.
\newblock \emph{\bibinfo{journal}{Nucleic Acids Research}}
  \textbf{\bibinfo{volume}{37}}, \bibinfo{pages}{D37--D40}
  (\bibinfo{year}{2008}).
\newblock \urlprefix\url{https://doi.org/10.1093/nar/gkn597}.

\bibitem{AAindex}
\bibinfo{author}{Kawashima, S.} \emph{et~al.}
\newblock \bibinfo{title}{{AAindex: amino acid index database, progress report
  2008}}.
\newblock \emph{\bibinfo{journal}{Nucleic Acids Research}}
  \textbf{\bibinfo{volume}{36}}, \bibinfo{pages}{D202--D205}
  (\bibinfo{year}{2007}).
\newblock \urlprefix\url{https://doi.org/10.1093/nar/gkm998}.

\bibitem{gkmSVM}
\bibinfo{author}{Ghandi, M.} \emph{et~al.}
\newblock \bibinfo{title}{{gkmSVM: an R package for gapped-kmer SVM}}.
\newblock \emph{\bibinfo{journal}{Bioinformatics}}
  \textbf{\bibinfo{volume}{32}}, \bibinfo{pages}{2205--2207}
  (\bibinfo{year}{2016}).
\newblock \urlprefix\url{https://doi.org/10.1093/bioinformatics/btw203}.

\bibitem{iLearnPlus}
\bibinfo{author}{Chen, Z.} \emph{et~al.}
\newblock \bibinfo{title}{{iLearnPlus: a comprehensive and automated
  machine-learning platform for nucleic acid and protein sequence analysis,
  prediction and visualization}}.
\newblock \emph{\bibinfo{journal}{Nucleic Acids Research}}
  \textbf{\bibinfo{volume}{49}}, \bibinfo{pages}{e60--e60}
  (\bibinfo{year}{2021}).
\newblock \urlprefix\url{https://doi.org/10.1093/nar/gkab122}.

\bibitem{BioSeq_Analysis2.0}
\bibinfo{author}{Liu, B.}, \bibinfo{author}{Gao, X.} \& \bibinfo{author}{Zhang,
  H.}
\newblock \bibinfo{title}{{BioSeq-Analysis2.0: an updated platform for
  analyzing DNA, RNA and protein sequences at sequence level and residue level
  based on machine learning approaches}}.
\newblock \emph{\bibinfo{journal}{Nucleic Acids Research}}
  \textbf{\bibinfo{volume}{47}}, \bibinfo{pages}{e127--e127}
  (\bibinfo{year}{2019}).
\newblock \urlprefix\url{https://doi.org/10.1093/nar/gkz740}.

\bibitem{BioSeq-BLM}
\bibinfo{author}{Li, H.-L.}, \bibinfo{author}{Pang, Y.-H.} \&
  \bibinfo{author}{Liu, B.}
\newblock \bibinfo{title}{{BioSeq-BLM: a platform for analyzing DNA, RNA and
  protein sequences based on biological language models}}.
\newblock \emph{\bibinfo{journal}{Nucleic Acids Research}}
  \textbf{\bibinfo{volume}{49}}, \bibinfo{pages}{e129--e129}
  (\bibinfo{year}{2021}).
\newblock \urlprefix\url{https://doi.org/10.1093/nar/gkab829}.

\bibitem{10.1371/journal.pcbi.1003711}
\bibinfo{author}{Ghandi, M.}, \bibinfo{author}{Lee, D.},
  \bibinfo{author}{Mohammad-Noori, M.} \& \bibinfo{author}{Beer, M.~A.}
\newblock \bibinfo{title}{Enhanced regulatory sequence prediction using gapped
  k-mer features}.
\newblock \emph{\bibinfo{journal}{PLOS Computational Biology}}
  \textbf{\bibinfo{volume}{10}}, \bibinfo{pages}{1--15} (\bibinfo{year}{2014}).
\newblock \urlprefix\url{https://doi.org/10.1371/journal.pcbi.1003711}.

\bibitem{lee2015method}
\bibinfo{author}{Lee, D.} \emph{et~al.}
\newblock \bibinfo{title}{A method to predict the impact of regulatory variants
  from dna sequence}.
\newblock \emph{\bibinfo{journal}{Nature genetics}}
  \textbf{\bibinfo{volume}{47}}, \bibinfo{pages}{955--961}
  (\bibinfo{year}{2015}).
\newblock \urlprefix\url{https://doi.org/10.1038/ng.3331}.

\bibitem{AlphaFold}
\bibinfo{author}{Jumper, J.} \emph{et~al.}
\newblock \bibinfo{title}{Highly accurate protein structure prediction with
  alphafold}.
\newblock \emph{\bibinfo{journal}{Nature}} \textbf{\bibinfo{volume}{596}},
  \bibinfo{pages}{583--589} (\bibinfo{year}{2021}).
\newblock \urlprefix\url{https://doi.org/10.1038/s41586-021-03819-2}.

\bibitem{Kipoi}
\bibinfo{author}{Avsec, {\v{Z}}.} \emph{et~al.}
\newblock \bibinfo{title}{The kipoi repository accelerates community exchange
  and reuse of predictive models for genomics}.
\newblock \emph{\bibinfo{journal}{Nature biotechnology}}
  \textbf{\bibinfo{volume}{37}}, \bibinfo{pages}{592--600}
  (\bibinfo{year}{2019}).
\newblock \urlprefix\url{https://doi.org/10.1038/s41587-019-0140-0}.

\bibitem{pysster}
\bibinfo{author}{Budach, S.} \& \bibinfo{author}{Marsico, A.}
\newblock \bibinfo{title}{{pysster: classification of biological sequences by
  learning sequence and structure motifs with convolutional neural networks}}.
\newblock \emph{\bibinfo{journal}{Bioinformatics}}
  \textbf{\bibinfo{volume}{34}}, \bibinfo{pages}{3035--3037}
  (\bibinfo{year}{2018}).
\newblock \urlprefix\url{https://doi.org/10.1093/bioinformatics/bty222}.

\bibitem{Selene}
\bibinfo{author}{Chen, K.~M.}, \bibinfo{author}{Cofer, E.~M.},
  \bibinfo{author}{Zhou, J.} \& \bibinfo{author}{Troyanskaya, O.~G.}
\newblock \bibinfo{title}{Selene: a pytorch-based deep learning library for
  sequence data}.
\newblock \emph{\bibinfo{journal}{Nature methods}}
  \textbf{\bibinfo{volume}{16}}, \bibinfo{pages}{315--318}
  (\bibinfo{year}{2019}).
\newblock \urlprefix\url{https://doi.org/10.1038/s41592-019-0360-8}.

\bibitem{DNABERT}
\bibinfo{author}{Ji, Y.}, \bibinfo{author}{Zhou, Z.}, \bibinfo{author}{Liu, H.}
  \& \bibinfo{author}{Davuluri, R.~V.}
\newblock \bibinfo{title}{{DNABERT: pre-trained Bidirectional Encoder
  Representations from Transformers model for DNA-language in genome}}.
\newblock \emph{\bibinfo{journal}{Bioinformatics}}
  \textbf{\bibinfo{volume}{37}}, \bibinfo{pages}{2112--2120}
  (\bibinfo{year}{2021}).
\newblock \urlprefix\url{https://doi.org/10.1093/bioinformatics/btab083}.

\bibitem{chen2016learning}
\bibinfo{author}{Chen, L.}, \bibinfo{author}{Cai, C.}, \bibinfo{author}{Chen,
  V.} \& \bibinfo{author}{Lu, X.}
\newblock \bibinfo{title}{Learning a hierarchical representation of the yeast
  transcriptomic machinery using an autoencoder model}.
\newblock \emph{\bibinfo{journal}{BMC Bioinformatics}}
  \textbf{\bibinfo{volume}{17}}, \bibinfo{pages}{S9} (\bibinfo{year}{2016}).
\newblock \urlprefix\url{https://doi.org/10.1186/s12859-015-0852-1}.

\bibitem{10.1093/bioinformatics/btw427}
\bibinfo{author}{Singh, R.}, \bibinfo{author}{Lanchantin, J.},
  \bibinfo{author}{Robins, G.} \& \bibinfo{author}{Qi, Y.}
\newblock \bibinfo{title}{{DeepChrome: deep-learning for predicting gene
  expression from histone modifications}}.
\newblock \emph{\bibinfo{journal}{Bioinformatics}}
  \textbf{\bibinfo{volume}{32}}, \bibinfo{pages}{i639--i648}
  (\bibinfo{year}{2016}).
\newblock \urlprefix\url{https://doi.org/10.1093/bioinformatics/btw427}.

\bibitem{10.1093/bioinformatics/btw255}
\bibinfo{author}{Zeng, H.}, \bibinfo{author}{Edwards, M.~D.},
  \bibinfo{author}{Liu, G.} \& \bibinfo{author}{Gifford, D.~K.}
\newblock \bibinfo{title}{{Convolutional neural network architectures for
  predicting DNA–protein binding}}.
\newblock \emph{\bibinfo{journal}{Bioinformatics}}
  \textbf{\bibinfo{volume}{32}}, \bibinfo{pages}{i121--i127}
  (\bibinfo{year}{2016}).
\newblock \urlprefix\url{https://doi.org/10.1093/bioinformatics/btw255}.

\bibitem{10.1093/nar/gkx177}
\bibinfo{author}{Zeng, H.} \& \bibinfo{author}{Gifford, D.~K.}
\newblock \bibinfo{title}{{Predicting the impact of non-coding variants on DNA
  methylation}}.
\newblock \emph{\bibinfo{journal}{Nucleic Acids Research}}
  \textbf{\bibinfo{volume}{45}}, \bibinfo{pages}{e99--e99}
  (\bibinfo{year}{2017}).
\newblock \urlprefix\url{https://doi.org/10.1093/nar/gkx177}.

\bibitem{7822593}
\bibinfo{author}{Min, X.}, \bibinfo{author}{Chen, N.}, \bibinfo{author}{Chen,
  T.} \& \bibinfo{author}{Jiang, R.}
\newblock \bibinfo{editor}{.} (ed.) \emph{\bibinfo{title}{Deepenhancer:
  Predicting enhancers by convolutional neural networks}}.
\newblock (ed.\bibinfo{editor}{.}) \emph{\bibinfo{booktitle}{2016 IEEE
  International Conference on Bioinformatics and Biomedicine (BIBM)}},
  \bibinfo{pages}{637--644} (\bibinfo{year}{2016}).
\newblock \urlprefix\url{https://doi.org/10.1109/BIBM.2016.7822593}.

\bibitem{10.1093/bioinformatics/bty228}
\bibinfo{author}{Aoki, G.} \& \bibinfo{author}{Sakakibara, Y.}
\newblock \bibinfo{title}{{Convolutional neural networks for classification of
  alignments of non-coding RNA sequences}}.
\newblock \emph{\bibinfo{journal}{Bioinformatics}}
  \textbf{\bibinfo{volume}{34}}, \bibinfo{pages}{i237--i244}
  (\bibinfo{year}{2018}).
\newblock \urlprefix\url{https://doi.org/10.1093/bioinformatics/bty228}.

\bibitem{zhou2015predicting}
\bibinfo{author}{Zhou, J.} \& \bibinfo{author}{Troyanskaya, O.~G.}
\newblock \bibinfo{title}{Predicting effects of noncoding variants with deep
  learning--based sequence model}.
\newblock \emph{\bibinfo{journal}{Nature methods}}
  \textbf{\bibinfo{volume}{12}}, \bibinfo{pages}{931--934}
  (\bibinfo{year}{2015}).
\newblock \urlprefix\url{https://doi.org/10.1038/nmeth.3547}.

\bibitem{alipanahi2015predicting}
\bibinfo{author}{Alipanahi, B.}, \bibinfo{author}{Delong, A.},
  \bibinfo{author}{Weirauch, M.~T.} \& \bibinfo{author}{Frey, B.~J.}
\newblock \bibinfo{title}{Predicting the sequence specificities of dna-and
  rna-binding proteins by deep learning}.
\newblock \emph{\bibinfo{journal}{Nature biotechnology}}
  \textbf{\bibinfo{volume}{33}}, \bibinfo{pages}{831--838}
  (\bibinfo{year}{2015}).
\newblock \urlprefix\url{https://doi.org/10.1038/nbt.3300}.

\bibitem{angermueller2017deepcpg}
\bibinfo{author}{Angermueller, C.}, \bibinfo{author}{Lee, H.~J.},
  \bibinfo{author}{Reik, W.} \& \bibinfo{author}{Stegle, O.}
\newblock \bibinfo{title}{Deepcpg: accurate prediction of single-cell dna
  methylation states using deep learning}.
\newblock \emph{\bibinfo{journal}{Genome biology}}
  \textbf{\bibinfo{volume}{18}}, \bibinfo{pages}{1--13} (\bibinfo{year}{2017}).

\bibitem{10.1093/bioinformatics/btx234}
\bibinfo{author}{Min, X.}, \bibinfo{author}{Zeng, W.}, \bibinfo{author}{Chen,
  N.}, \bibinfo{author}{Chen, T.} \& \bibinfo{author}{Jiang, R.}
\newblock \bibinfo{title}{{Chromatin accessibility prediction via convolutional
  long short-term memory networks with k-mer embedding}}.
\newblock \emph{\bibinfo{journal}{Bioinformatics}}
  \textbf{\bibinfo{volume}{33}}, \bibinfo{pages}{i92--i101}
  (\bibinfo{year}{2017}).
\newblock \urlprefix\url{https://doi.org/10.1093/bioinformatics/btx234}.

\bibitem{10.1093/nar/gkw226}
\bibinfo{author}{Quang, D.} \& \bibinfo{author}{Xie, X.}
\newblock \bibinfo{title}{{DanQ: a hybrid convolutional and recurrent deep
  neural network for quantifying the function of DNA sequences}}.
\newblock \emph{\bibinfo{journal}{Nucleic Acids Research}}
  \textbf{\bibinfo{volume}{44}}, \bibinfo{pages}{e107--e107}
  (\bibinfo{year}{2016}).
\newblock \urlprefix\url{https://doi.org/10.1093/nar/gkw226}.

\bibitem{Dataset1}
\bibinfo{author}{Lu, W.} \emph{et~al.}
\newblock \bibinfo{title}{Use chou's 5-step rule to predict dna-binding
  proteins with evolutionary information}.
\newblock \emph{\bibinfo{journal}{BioMed Research International}}
  \textbf{\bibinfo{volume}{2020}}, \bibinfo{pages}{6984045}
  (\bibinfo{year}{2020}).
\newblock \urlprefix\url{https://doi.org/10.1155/2020/6984045}.

\bibitem{Dataset2}
\bibinfo{author}{Hui, X.} \emph{et~al.}
\newblock \bibinfo{title}{T3sepp: an integrated prediction pipeline for
  bacterial type iii secreted effectors}.
\newblock \emph{\bibinfo{journal}{mSystems}} \textbf{\bibinfo{volume}{5}},
  \bibinfo{pages}{e00288--20} (\bibinfo{year}{2020}).
\newblock
  \urlprefix\url{https://journals.asm.org/doi/abs/10.1128/mSystems.00288-20}.

\bibitem{Dataset3}
\bibinfo{author}{Meng, C.}, \bibinfo{author}{Zhang, J.}, \bibinfo{author}{Ye,
  X.}, \bibinfo{author}{Guo, F.} \& \bibinfo{author}{Zou, Q.}
\newblock \bibinfo{title}{Review and comparative analysis of machine
  learning-based phage virion protein identification methods}.
\newblock \emph{\bibinfo{journal}{Biochimica et Biophysica Acta (BBA) -
  Proteins and Proteomics}} \textbf{\bibinfo{volume}{1868}},
  \bibinfo{pages}{140406} (\bibinfo{year}{2020}).
\newblock
  \urlprefix\url{https://www.sciencedirect.com/science/article/pii/S1570963920300479}.

\bibitem{Dataset4}
\bibinfo{author}{Barukab, O.}, \bibinfo{author}{Khan, Y.~D.},
  \bibinfo{author}{Khan, S.~A.} \& \bibinfo{author}{Chou, K.-C.}
\newblock \bibinfo{title}{isulfotyr-pseaac: Identify tyrosine sulfation sites
  by incorporating statistical moments via chou’s 5-steps rule and pseudo
  components}.
\newblock \emph{\bibinfo{journal}{Current Genomics}}
  \textbf{\bibinfo{volume}{20}}, \bibinfo{pages}{306--320}
  (\bibinfo{year}{2019}).
\newblock \urlprefix\url{https://doi.org/10.2174/1389202920666190819091609}.

\bibitem{Dataset5}
\bibinfo{author}{Li, T.}, \bibinfo{author}{Song, R.}, \bibinfo{author}{Yin,
  Q.}, \bibinfo{author}{Gao, M.} \& \bibinfo{author}{Chen, Y.}
\newblock \bibinfo{title}{Identification of s-nitrosylation sites based on
  multiple features combination}.
\newblock \emph{\bibinfo{journal}{Scientific Reports}}
  \textbf{\bibinfo{volume}{9}}, \bibinfo{pages}{3098} (\bibinfo{year}{2019}).
\newblock \urlprefix\url{https://doi.org/10.1038/s41598-019-39743-9}.

\bibitem{Dataset6}
\bibinfo{author}{Dou, L.}, \bibinfo{author}{Li, X.}, \bibinfo{author}{Zhang,
  L.}, \bibinfo{author}{Xiang, H.} \& \bibinfo{author}{Xu, L.}
\newblock \bibinfo{title}{iglu\_adaboost: Identification of lysine
  glutarylation using the adaboost classifier}.
\newblock \emph{\bibinfo{journal}{Journal of Proteome Research}}
  \textbf{\bibinfo{volume}{20}}, \bibinfo{pages}{191--201}
  (\bibinfo{year}{2021}).
\newblock \urlprefix\url{https://doi.org/10.1021/acs.jproteome.0c00314}.
\newblock \bibinfo{note}{PMID: 33090794}.

\bibitem{Dataset7to10}
\bibinfo{author}{Jia, J.}, \bibinfo{author}{Liu, Z.}, \bibinfo{author}{Xiao,
  X.}, \bibinfo{author}{Liu, B.} \& \bibinfo{author}{Chou, K.-C.}
\newblock \bibinfo{title}{icar-psecp: identify carbonylation sites in proteins
  by monte carlo sampling and incorporating sequence coupled effects into
  general pseaac}.
\newblock \emph{\bibinfo{journal}{Oncotarget}} \textbf{\bibinfo{volume}{7}},
  \bibinfo{pages}{34558} (\bibinfo{year}{2016}).
\newblock \urlprefix\url{https://doi.org/10.18632/oncotarget.9148}.

\bibitem{hsic}
\bibinfo{author}{Ding, Y.}, \bibinfo{author}{Tang, J.} \& \bibinfo{author}{Guo,
  F.}
\newblock \bibinfo{title}{Identification of drug–target interactions via dual
  laplacian regularized least squares with multiple kernel fusion}.
\newblock \emph{\bibinfo{journal}{Knowledge-Based Systems}}
  \textbf{\bibinfo{volume}{204}}, \bibinfo{pages}{106254}
  (\bibinfo{year}{2020}).
\newblock
  \urlprefix\url{https://www.sciencedirect.com/science/article/pii/S0950705120304494}.

\bibitem{hkam}
\bibinfo{author}{Wang, Y.}, \bibinfo{author}{Liu, X.}, \bibinfo{author}{Dou,
  Y.}, \bibinfo{author}{Lv, Q.} \& \bibinfo{author}{Lu, Y.}
\newblock \bibinfo{title}{Multiple kernel learning with hybrid kernel alignment
  maximization}.
\newblock \emph{\bibinfo{journal}{Pattern Recognition}}
  \textbf{\bibinfo{volume}{70}}, \bibinfo{pages}{104--111}
  (\bibinfo{year}{2017}).
\newblock
  \urlprefix\url{https://www.sciencedirect.com/science/article/pii/S0031320317301863}.

\bibitem{PhysicochemicalProperties1}
\bibinfo{author}{Richardson, J.~S.} \& \bibinfo{author}{Richardson, D.~C.}
\newblock \bibinfo{title}{Amino acid preferences for specific locations at the
  ends of \&\#x3b1; helices}.
\newblock \emph{\bibinfo{journal}{Science}} \textbf{\bibinfo{volume}{240}},
  \bibinfo{pages}{1648--1652} (\bibinfo{year}{1988}).
\newblock
  \urlprefix\url{https://www.science.org/doi/abs/10.1126/science.3381086}.

\bibitem{PhysicochemicalProperties2}
\bibinfo{author}{Meirovitch, H.}, \bibinfo{author}{Rackovsky, S.} \&
  \bibinfo{author}{Scheraga, H.~A.}
\newblock \bibinfo{title}{Empirical studies of hydrophobicity. 1. effect of
  protein size on the hydrophobic behavior of amino acids}.
\newblock \emph{\bibinfo{journal}{Macromolecules}}
  \textbf{\bibinfo{volume}{13}}, \bibinfo{pages}{1398--1405}
  (\bibinfo{year}{1980}).
\newblock \urlprefix\url{https://doi.org/10.1021/ma60078a013}.

\bibitem{PhysicochemicalProperties10}
\bibinfo{author}{Cornette, J.~L.} \emph{et~al.}
\newblock \bibinfo{title}{Hydrophobicity scales and computational techniques
  for detecting amphipathic structures in proteins}.
\newblock \emph{\bibinfo{journal}{Journal of Molecular Biology}}
  \textbf{\bibinfo{volume}{195}}, \bibinfo{pages}{659--685}
  (\bibinfo{year}{1987}).
\newblock
  \urlprefix\url{https://www.sciencedirect.com/science/article/pii/0022283687901896}.

\bibitem{PhysicochemicalProperties3}
\bibinfo{author}{Geisow, M.~J.} \& \bibinfo{author}{Roberts, R.~D.}
\newblock \bibinfo{title}{Amino acid preferences for secondary structure vary
  with protein class}.
\newblock \emph{\bibinfo{journal}{International Journal of Biological
  Macromolecules}} \textbf{\bibinfo{volume}{2}}, \bibinfo{pages}{387--389}
  (\bibinfo{year}{1980}).
\newblock
  \urlprefix\url{https://www.sciencedirect.com/science/article/pii/0141813080900239}.

\bibitem{PhysicochemicalProperties5}
\bibinfo{author}{Biou, V.}, \bibinfo{author}{Gibrat, J.},
  \bibinfo{author}{Levin, J.}, \bibinfo{author}{Robson, B.} \&
  \bibinfo{author}{Garnier, J.}
\newblock \bibinfo{title}{{Secondary structure prediction: combination of three
  different methods}}.
\newblock \emph{\bibinfo{journal}{Protein Engineering, Design and Selection}}
  \textbf{\bibinfo{volume}{2}}, \bibinfo{pages}{185--191}
  (\bibinfo{year}{1988}).
\newblock \urlprefix\url{https://doi.org/10.1093/protein/2.3.185}.

\bibitem{PhysicochemicalProperties4}
\bibinfo{author}{Oobatake, M.} \& \bibinfo{author}{Ooi, T.}
\newblock \bibinfo{title}{An analysis of non-bonded energy of proteins}.
\newblock \emph{\bibinfo{journal}{Journal of Theoretical Biology}}
  \textbf{\bibinfo{volume}{67}}, \bibinfo{pages}{567--584}
  (\bibinfo{year}{1977}).
\newblock
  \urlprefix\url{https://www.sciencedirect.com/science/article/pii/0022519377900583}.

\bibitem{PhysicochemicalProperties6}
\bibinfo{author}{Nakashima, H.} \& \bibinfo{author}{Nishikawa, K.}
\newblock \bibinfo{title}{The amino acid composition is different between the
  cytoplasmic and extracellular sides in membrane proteins}.
\newblock \emph{\bibinfo{journal}{FEBS Letters}}
  \textbf{\bibinfo{volume}{303}}, \bibinfo{pages}{141--146}
  (\bibinfo{year}{1992}).
\newblock
  \urlprefix\url{https://www.sciencedirect.com/science/article/pii/001457939280506C}.

\bibitem{PhysicochemicalProperties7}
\bibinfo{author}{Zimmerman, J.}, \bibinfo{author}{Eliezer, N.} \&
  \bibinfo{author}{Simha, R.}
\newblock \bibinfo{title}{The characterization of amino acid sequences in
  proteins by statistical methods}.
\newblock \emph{\bibinfo{journal}{Journal of Theoretical Biology}}
  \textbf{\bibinfo{volume}{21}}, \bibinfo{pages}{170--201}
  (\bibinfo{year}{1968}).
\newblock
  \urlprefix\url{https://www.sciencedirect.com/science/article/pii/0022519368900696}.

\bibitem{PhysicochemicalProperties8}
\bibinfo{author}{Sneath, P.}
\newblock \bibinfo{title}{Relations between chemical structure and biological
  activity in peptides}.
\newblock \emph{\bibinfo{journal}{Journal of Theoretical Biology}}
  \textbf{\bibinfo{volume}{12}}, \bibinfo{pages}{157--195}
  (\bibinfo{year}{1966}).
\newblock
  \urlprefix\url{https://www.sciencedirect.com/science/article/pii/0022519366901123}.

\bibitem{PhysicochemicalProperties9}
\bibinfo{author}{Radzicka, A.} \& \bibinfo{author}{Wolfenden, R.}
\newblock \bibinfo{title}{Comparing the polarities of the amino acids:
  side-chain distribution coefficients between the vapor phase, cyclohexane,
  1-octanol, and neutral aqueous solution}.
\newblock \emph{\bibinfo{journal}{Biochemistry}} \textbf{\bibinfo{volume}{27}},
  \bibinfo{pages}{1664--1670} (\bibinfo{year}{1988}).
\newblock \urlprefix\url{https://doi.org/10.1021/bi00405a042}.

\bibitem{PhysicochemicalProperties11}
\bibinfo{author}{Guy, H.}
\newblock \bibinfo{title}{Amino acid side-chain partition energies and
  distribution of residues in soluble proteins}.
\newblock \emph{\bibinfo{journal}{Biophysical Journal}}
  \textbf{\bibinfo{volume}{47}}, \bibinfo{pages}{61--70}
  (\bibinfo{year}{1985}).
\newblock
  \urlprefix\url{https://www.sciencedirect.com/science/article/pii/S0006349585838777}.

\bibitem{PhysicochemicalProperties12}
\bibinfo{author}{Yang, J.-M.} \emph{et~al.}
\newblock \bibinfo{title}{Gem: A gaussian evolutionary method for predicting
  protein side-chain conformations}.
\newblock \emph{\bibinfo{journal}{Protein Science}}
  \textbf{\bibinfo{volume}{11}}, \bibinfo{pages}{1897--1907}
  (\bibinfo{year}{2002}).
\newblock
  \urlprefix\url{https://onlinelibrary.wiley.com/doi/abs/10.1110/ps.4940102}.

\end{thebibliography}

\section*{Acknowledgements}

This work is supported by the National Natural Science Foundation of China (NSFC 62250028, 62172076, U22A2038), Zhejiang Provincial Natural Science Foundation of China (grant no. LY23F020003), the Municipal Government of Quzhou (grant no. 2022D040).

\section*{Author contributions}

Yizheng Wang did the experiments and wrote the manuscript. Yizheng Wang, Yixiao Zhai, Yijie Ding, and Quan Zou designed the method. Yizheng Wang, Yixiao Zhai, Yijie Ding, and Quan Zou revised the manuscript. All authors have read and approved the final manuscript.

\section*{Competing interests}

The authors declare no competing interests.

\begin{appendices}

\section{Supplementary material}\label{Supplementary material}

\begin{table}[!ht]
\caption{Summary of processed physicochemical properties of amino acids.}\label{Removed data}%
\begin{tabular}{lcccccccccc}
\toprule
& SC1\footnotemark[1]  & SC2\footnotemark[2]\\
Amino acid & $P_{11}$ & $P_{12}$ \\
\midrule
Ala (A) & 0.54 & NA \\
Arg (R) & -0.16 & 0.62 \\
Asn (N) & 0.38 & 0.76 \\
Asp (D) & 0.65 & 0.66 \\
Cys (C) & -1.13 & 0.83 \\
Gln (Q) & 0.05 & 0.59 \\
Glu (E) & 0.38 & 0.73 \\
Gly (G) & NA & NA \\
His (H) & -0.59 & 0.92 \\
Ile (I) & -2 & 0.88 \\
Leu (L) & -1.08 & 0.89 \\
Lys (K) & 0.48 & 0.77 \\
Met (M) & -0.97 & 0.77 \\
Phe (F) & -1.51 & 0.92 \\
Pro (P) & -0.22 & 0.94 \\
Ser (S) & 0.65 & 0.58 \\
Thr (T) & 0.27 & 0.73 \\
Trp (W) & -1.61 & 0.86 \\
Tyr (Y) & -1.13 & 0.93 \\
Val (V) & -0.75 & 0.88 \\
\botrule
\end{tabular}
\footnotetext[1]{The dataset is obtained from the study\cite{PhysicochemicalProperties11}.}
\footnotetext[2]{The dataset is obtained from the study\cite{PhysicochemicalProperties12}.}
\end{table}

\begin{table}[!ht]
\caption{Dictionaries for grouping $D_{1}$.}\label{D1}%
\begin{tabular}{cccccccc}
\toprule
Dictionary & Group & Amino Acid\\
\midrule
\multirow{7}{*}{$D_1$} 
& {$G_1$} & A, W \\
& {$G_2$} & D \\
& {$G_3$} & R, Q, G, F, S \\
& {$G_4$} & E \\
& {$G_5$} & L, Y \\
& {$G_6$} & N, H, I, P, T, V \\
& {$G_7$} & C, K, M \\
\botrule
\end{tabular}
\end{table}

\begin{table}[!ht]
\caption{Dictionaries for grouping $D_{2}$.}\label{D2}%
\begin{tabular}{cccccccc}
\toprule
Dictionary & Group & Amino Acid\\
\midrule
\multirow{7}{*}{$D_2$} 
& {$G_1$} & R, S, T, Y \\
& {$G_2$} & C, I, L, M, F \\
& {$G_3$} & Q, K \\
& {$G_4$} & P \\
& {$G_5$} & W, V \\
& {$G_6$} & N, D, E \\
& {$G_7$} & A, G, H \\
\botrule
\end{tabular}
\end{table}

\begin{table}[!ht]
\caption{Dictionaries for grouping $D_{3}$.}\label{D3}%
\begin{tabular}{cccccccc}
\toprule
Dictionary & Group & Amino Acid\\
\midrule
\multirow{7}{*}{$D_3$} 
& {$G_1$} & D, Q, Y \\
& {$G_2$} & A, I, L, T, V \\
& {$G_3$} & N, F, S \\
& {$G_4$} & G, P \\
& {$G_5$} & C, W \\
& {$G_6$} & E, M \\
& {$G_7$} & R, H, K \\
\botrule
\end{tabular}
\end{table}

\begin{table}[!ht]
\caption{Grouping for amino acid $D_{4}$.}\label{D4}%
\begin{tabular}{cccccccc}
\toprule
Dictionary & Group & Amino Acid\\
\midrule
\multirow{7}{*}{$D_4$} 
& {$G_1$} & R, G, H, T \\
& {$G_2$} & A, P, S \\
& {$G_3$} & W \\
& {$G_4$} & D, K \\
& {$G_5$} & C, L, M, Y \\
& {$G_6$} & I, F, V \\
& {$G_7$} & N, Q, E \\
\botrule
\end{tabular}
\end{table}

\begin{table}[!ht]
\caption{Grouping for amino acid $D_{5}$.}\label{D5}%
\begin{tabular}{cccccccc}
\toprule
Dictionary & Group & Amino Acid\\
\midrule
\multirow{6}{*}{$D_5$} 
& {$G_1$} & H, Y \\
& {$G_2$} & A, G, P, T \\
& {$G_3$} & K \\
& {$G_4$} & I, L, M, W, V \\
& {$G_5$} & R, N, D, Q, E, S \\
& {$G_6$} & C, F \\
\botrule
\end{tabular}
\end{table}

\begin{table}[!ht]
\caption{Grouping for amino acid $D_{6}$.}\label{D6}%
\begin{tabular}{cccccccc}
\toprule
Dictionary & Group & Amino Acid\\
\midrule
\multirow{6}{*}{$D_6$} 
& {$G_1$} & G, S, T \\
& {$G_2$} & N, C, M, P, W, Y \\
& {$G_3$} & L \\
& {$G_4$} & I, V \\
& {$G_5$} & A, F \\
& {$G_6$} & R, D, Q, E, H, K \\
\botrule
\end{tabular}
\end{table}

\begin{table}[!ht]
\caption{Grouping for amino acid $D_{7}$.}\label{D7}%
\begin{tabular}{cccccccc}
\toprule
Dictionary & Group & Amino Acid\\
\midrule
\multirow{7}{*}{$D_7$} 
& {$G_1$} & M, P, Y, V \\
& {$G_2$} & F \\
& {$G_3$} & H, S \\
& {$G_4$} & R, N, G \\
& {$G_5$} & D, C, E, K \\
& {$G_6$} & A, Q, T \\
& {$G_7$} & I, L, W \\
\botrule
\end{tabular}
\end{table}

\begin{table}[!ht]
\caption{Grouping for amino acid $D_{8}$.}\label{D8}%
\begin{tabular}{cccccccc}
\toprule
Dictionary & Group & Amino Acid\\
\midrule
\multirow{7}{*}{$D_8$} 
& {$G_1$} & A, G, P \\
& {$G_2$} & C, I, L, T \\
& {$G_3$} & D, E \\
& {$G_4$} & R, H, K, M, Y \\
& {$G_5$} & F, S, W \\
& {$G_6$} & V \\
& {$G_7$} & N, Q \\
\botrule
\end{tabular}
\end{table}

\begin{table}[!ht]
\caption{Grouping for amino acid $D_{9}$.}\label{D9}%
\begin{tabular}{cccccccc}
\toprule
Dictionary & Group & Amino Acid\\
\midrule
\multirow{7}{*}{$D_9$} 
& {$G_1$} & C, I, L, M, F \\
& {$G_2$} & K \\
& {$G_3$} & N, G, S, T \\
& {$G_4$} & W, V \\
& {$G_5$} & H, Y \\
& {$G_6$} & R, D, Q, E \\
& {$G_7$} & A, P \\
\botrule
\end{tabular}
\end{table}

\begin{table}[!ht]
\caption{Grouping for amino acid $D_{10}$.}\label{D10}%
\begin{tabular}{cccccccc}
\toprule
Dictionary & Group & Amino Acid\\
\midrule
\multirow{7}{*}{$D_{10}$} 
& {$G_1$} & L \\
& {$G_2$} & A, N, G, P, S \\
& {$G_3$} & D \\
& {$G_4$} & I, M, F, V \\
& {$G_5$} & C, Y \\
& {$G_6$} & Q, E, K, T \\
& {$G_7$} & R, H, W \\
\botrule
\end{tabular}
\end{table}

\end{appendices}

\end{document}